\documentclass[%
 reprint,
 superscriptaddress,
 noeprint,
 nofootinbib,
 amsmath,amssymb,
 aps,
 prx,
floatfix,
]{revtex4-2}
\usepackage[svgnames]{xcolor}
\usepackage{amsfonts}
\usepackage{mathtools} % for mathclap
\usepackage{braket}
\usepackage{hyperref}
\usepackage{siunitx}
\usepackage{graphicx}
\usepackage[ruled]{algorithm2e}
\usepackage{booktabs}
\usepackage{array}
\usepackage{listings}
\usepackage{tikz}
\usepackage[capitalise]{cleveref} % should be loaded last, pretty much always

\graphicspath{%
    {figures/validation/},
    {figures/compressibility/},
    {figures/tensor_to_mps/},
    {figures/heat_release/},
    {figures/left_shift_fsm/},
    {figures/left_and_right_shift/},
    {figures/shocklets/},
    {figures/high_damkohler/},
    {figures/reynolds_scaling/},
    {figures/infidelity_threshold/},
    {figures/single_schmidt_example},
    {figures/TDJ_profile},
    {figures/index_ordering}
}

\newcommand{\pfrac}[2]{\ensuremath{\frac{\partial#1}{\partial#2}}}
\newcommand{\Pe}{\ensuremath{\mathrm{Pe}}}
\newcommand{\Da}{\ensuremath{\mathrm{Da}}}

\newcommand{\Ma}{\ensuremath{\mathrm{Ma}_o}}
\newcommand{\MA}{\ensuremath{\mathrm{Ma}_o^2}}
\newcommand{\Maa}{\ensuremath{\mathrm{Ma}}}
\newcommand{\Rey}{\ensuremath{\mathrm{Re}}}
\newcommand{\Peclet}{P\'eclet}
\newcommand{\Damkohler}{Damk\"ohler}

\let\oldsim\sim
\renewcommand{\sim}{{\oldsim}}

\newcommand{\LS}{\ensuremath{\operatorname{LS}}}
\newcommand{\RS}{\ensuremath{\operatorname{RS}}}
\newcommand{\dof}{\ensuremath{\operatorname{dof}}}
\newcommand{\ord}[1]{\ensuremath{\mathcal{O}(#1)}}

\newcommand{\mpo}[2]{O_{#1}^{#2}}
\newcommand{\res}[2]{B_{#1}^{#2}}
\newcommand{\dx}{\hat{\partial}_x}
\newcommand{\dy}{\hat{\partial}_y}

\newcommand{\xx}{\ensuremath{x}}
\newcommand{\yy}{\ensuremath{y}}
\newcommand{\II}{\ensuremath{I}}
\newcommand{\YY}{\ensuremath{Y}}

\newcommand{\fC}[1]{\ensuremath{f_C}}
\newcommand{\fF}[1]{\ensuremath{f_F}}
\DeclareMathOperator*{\argmin}{arg\,min}

\DeclareMathOperator*{\SIZEOF}{size}

\newcommand{\mT}{\ensuremath{\mathcal{T}}}
\newcommand{\chicompete}{\ensuremath{\chi_c}}
\newcommand{\chimax}{\ensuremath{r}}

\newcommand{\mpsA}{\ensuremath{\ket{\psi_A}}}
\newcommand{\mpsB}{\ensuremath{\ket{\psi_B}}}
\newcommand{\mpsC}{\ensuremath{\ket{\psi_C}}}
\newcommand{\psiDNS}{\ensuremath{\psi_{\text{DNS}}}}
\newcommand{\psiMPS}{\ensuremath{\psi_{\text{MPS}}}}

\newcommand{\sizeof}[1]{\ensuremath{\SIZEOF\left(#1\right)}}

\newcommand{\MPO}{\ensuremath{\widehat{G}}}

\DeclareRobustCommand{\rchi}{{\mathpalette\irchi\relax}}
\newcommand{\irchi}[2]{\raisebox{\depth}{$#1\chi$}} % inner command, used by \rchi

\crefname{section}{Sec.}{Secs.}

\begin{document}
\title{Matrix Product State Simulation of Reacting Shear Flows}
\author{Robert Pinkston}
\affiliation{Mechanical Engineering and Materials Science, University of Pittsburgh, Pittsburgh, PA 15261, USA}
\author{Nikita Gourianov}
\affiliation{Mechanical Engineering and Materials Science, University of Pittsburgh, Pittsburgh, PA 15261, USA}
\affiliation{Clarendon Laboratory, University of Oxford, Oxford OX13PU, UK}
\affiliation{Datalogisk Institut, University of Copenhagen, DK}
\author{Hirad Alipanah}
\affiliation{Mechanical Engineering and Materials Science, University of Pittsburgh, Pittsburgh, PA 15261, USA}
\author{Peyman Givi}
\affiliation{Mechanical Engineering and Materials Science, University of Pittsburgh, Pittsburgh, PA 15261, USA}
\affiliation{Petroleum  Engineering, University of Pittsburgh, Pittsburgh, PA 15261, USA}
\author{Dieter Jaksch}
\affiliation{Clarendon Laboratory, University of Oxford, Oxford OX13PU, UK}
\affiliation{Institut f\"{u}r Quantenphysik, Universit\"{a}t Hamburg, Luruper Chaussee 149, 22761 Hamburg, Germany}
\author{Juan Jos\'{e} Mendoza-Arenas}
\affiliation{Mechanical Engineering and Materials Science, University of Pittsburgh, Pittsburgh, PA 15261, USA}
\affiliation{Physics and Astronomy, University of Pittsburgh, Pittsburgh, PA 15261, USA}

\date{\today}

\begin{abstract}
    Direct numerical simulation (DNS) of turbulent reactive flows has been the subject of significant research interest for several  decades.
    Accurate prediction of the effects of turbulence on the rate of reactant conversion, and the subsequent influence of chemistry on hydrodynamics remain a challenge in combustion modeling.
    The key issue in DNS is to account for the wide range of temporal and spatial physical scales that are caused by complex interactions of turbulence and chemistry.
    In this work, a new computational methodology  is developed that is shown to provide a viable alternative to DNS.
    The framework is the matrix product state (MPS), a form of tensor network (TN) as used in computational many body physics.
    The MPS is a well-established ansatz for efficiently representing many types of quantum states in condensed matter systems, allowing for an exponential compression of the required memory compared to exact diagonalization methods.
    Due to the success of MPS in quantum physics, the ansatz has been adapted to problems outside its historical domain, notably computational fluid dynamics.
    Here, the MPS is used for computational simulation of a shear flow under non-reacting and nonpremixed chemically reacting conditions.
    Reductions of 30\% in memory are demonstrated for all transport variables, accompanied by excellent agreements with DNS.
    The anastaz accurately captures all pertinent flow physics such as reduced mixing due to exothermicity \& compressibility, and the formation of eddy shocklets at high Mach numbers.
    \textit{A priori} analysis of DNS data at higher Reynolds numbers shows compressions as large as 99.99\% for some of the transport variables.
    This level of compression is encouraging and promotes the use of MPS for simulations of complex turbulent combustion systems.
\end{abstract}

\maketitle{}

\section{Introduction}\label{sec:intro}
Reacting shear flows describe a wide variety of important applications ranging from a simple household furnace to a solid rocket motor.
Such flows involve a complex interaction between the hydrodynamics, governed by the Navier-Stokes equations, and the chemistry.
The heat released by reactions couples the chemical kinetics with the mass, momentum, and energy equations and leads to  physically rich and interesting phenomena~\cite{masri2021Challenges,wang2016Numerical,livescu2020Modeling}.
Computational scaling represents a major obstacle in the simulation of reacting shear flows~\cite{pope2013Small}.
Such flows are parameterized by several non-dimensional numbers, the most relevant of which is the Reynolds number \Rey{}.
In turbulent flows, the number of computational grids required to fully resolve the flow fields scales approximately as \ord{(\Rey{}^{3/4})^\mathcal{N}} where \(\mathcal{N}\) is the dimensionality~\cite{coleman2010Primer} and applications of \(\Rey{} > \ord{\num{e5}}\) are not uncommon~\cite{neuhaus2020Generation,wood2015Reynolds,wood2012Review}.
Adding the chemistry further increases the computational challenge.
Even small flames can require on the order of billions of grid points for accurate simulations~\cite{domingo2023Recent}.

A similar scaling challenge is present in the simulation of many-body quantum systems such as lattice models for quantum materials~\cite{coleman2015Introduction}.
Lattices with \(N\) sites are described by wave functions that are linear combinations of \(d^N\) basis elements, where \(d\) is the dimension of the local Hilbert space.
The wavefunctions then require the \(d^N\) coefficients to be fully described, which represents a daunting exponential scaling to overcome.
This difficult scaling led to the development of tensor network (TN) methods, algorithms built to address exponential scaling on classic hardware~\cite{banuls2023Tensor,schollwoeck2011densitymatrix,orus2019Tensora,paeckel2019Timeevolutiona,orus2014practical,verstraete2008Matrixa}.
TN methods exploit that in weakly-correlated systems, only a small subset of the \ord{d^N} wavefunction coefficients is needed to accurately describe the important physical properties~\cite{hastings2007Area,wolf2008Area}.
For example, the matrix product state (MPS) TN has been successfully  used to calculate ground states and time evolution of one- and two-dimensional systems with gapped Hamiltonians and short range interactions~\cite{stoudenmire2012Studyinga,schollwoeck2011densitymatrix,hastings2007Area,hastings2006Solving,wolf2008Area,vidal2008Class,shi2006Classical}.
The MPS can achieve an exponential truncation of the data needed to describe the system and with very little error~\cite{vidal2003Entanglement}.
This impressive truncation has motivated disciplines outside of  many-body quantum physics to explore the TN ansatz to efficiently encode information~\cite{swingle2012Entanglement,oseledets2011TENSORTRAIN,cichocki2016Tensor,cichocki2017Tensor,stoudenmire2016Supervisedb,orus2019Quantum}, particularly within the domain of fluid mechanics~\cite{gourianov2022Exploiting,gourianov2022Quantum,kiffner2023Tensora,adak2024Tensorb,holscher2025Quantuminspired,gourianov2025Tensor,danis2025Tensortraina,hulst2025QuantumInspired,connor2025Tensor,kornev2023Numericalb,ghahremani2024Crossa,pisoni2025Compression,peddinti2024Quantuminspired,gomez-lozada2025Simulating}.

The objective of this study is to demonstrate a new method for accurately simulating reacting shear flows while simultaneously resolving computational scaling challenges.
A two-dimensional reacting flow, described by six coupled partial differential equations, is encoded and time evolved entirely in the MPS ansatz.
Importantly, the transport variables (velocity, temperature, species, etc.) and all of their spatial derivatives  (e.g., velocity gradients and shear stresses) are truncated throughout the simulation.
The impact of compressibility (characterized by the Mach number) and heat release is assessed in the truncated simulation.
The benefits of the TN ansatz are evaluated and show that compression ratios on the order of \ord{\num{e-5}} are theoretically possible at high \Rey{}.
This property of   the MPS is very effective in dealing with  the scaling challenges of turbulent reacting flows.

This paper is structured as follows.
The governing equations of compressible reacting flows, as well as the description of geometry, boundary conditions, initial conditions, and important metrics, are introduced in \cref{sec:gov}.
A brief introduction to the MPS ansatz, matrix product operators (MPOs), and the MPS time-evolution algorithm is given in \cref{sec:MPS}.
The algorithm validation, MPS truncation error, flow physics, and the effectiveness of the anstaz are discussed  in \cref{sec:results}.
\section{Governing Equations and Problem Description}\label{sec:gov}
Six equations are required to mathematical describe two-dimensional compressible flow involving two chemical  species~\cite{poinsot2005Theoretical}.
All equations are in non-dimensional form unless otherwise stated.
The non-dimensionalization parameters are given in \cref{app:non_dimen}\@.
The chemical reaction is a binary second order exothermic reaction between species with mass fractions \(c_1\) and \(c_2\), which results in a product with mass fraction \(c_3\).
The conservation of mass is given as
\begin{equation}
    \pfrac{\rho}{t} + \pfrac{\rho u_i}{x_i} = 0,\label{eqn:cons_of_mass}
\end{equation}
with \(\rho\) the mass density and \(u_i\) the velocity components (\(u\) and \(v\)).
The equations of the conservation of momentum for both coordinate directions are given by
\begin{align}
    \pfrac{\rho u_i}{t} + \pfrac{\rho u_j u_i}{x_j} = -\pfrac{p}{x_i} + \frac{1}{\Rey{}}\pfrac{\tau_{ij}}{x_j},\label{eqn:cons_of_momentum}
\end{align}
where \(p\) is the pressure and \(\tau_{ij}\) are the viscous shear stresses.
Conservation of energy is given as
\begin{equation}
    \pfrac{\rho e_T}{t} + \pfrac{\rho u_i e_T}{x_i} = - \pfrac{q_i}{x_i} + \frac{1}{\Rey{}}\pfrac{\tau_{ij} u_i}{x_j} - \pfrac{u_j p}{x_j},\label{eqn:cons_of_energy}
\end{equation}
where \(e_T =  |\vec{V}|^2 / 2 + e (1 + c_e c_3)\), \(|\vec{V}|^2 = u_i u_i\), \(e\) is the internal energy, \(c_e\) is the heat release parameter, and \(q_i\) are the components of the heat flux vector~\cite{givi1991Effects}.
The transport of species \(i\) is given by
\begin{equation}
    \pfrac{\rho c_i}{t} + \pfrac{}{x_j} \left(\rho u_j c_i - \frac{1}{\Pe{}}\pfrac{c_i}{x_j}\right) = \dot{\omega}_i,\label{eqn:species}
\end{equation}
where \Pe{} is the \Peclet{} number and \(\dot{\omega}_i\) is the reaction rate parameterized by the \Damkohler{} number \Da{}:
\begin{equation}
    \dot{\omega}_1 = \dot{\omega}_2 = -\Da{}\rho c_1 c_2.\label{eqn:rxn_rate}
\end{equation}
\Cref{eqn:cons_of_mass,eqn:cons_of_momentum,eqn:cons_of_energy,eqn:species} are closed by assuming a calorically perfect gas~\cite{anderson1995Computational}.
The equations are conveniently represented in vector form:
\begin{equation}
    \pfrac{\vec{U}}{t} = - \left(\pfrac{\vec{F}}{x} + \pfrac{\vec{G}}{y}\right) + \vec{S}.\label{eqn:gov}
\end{equation}
Here \(\vec{U}\) is the solution vector, \(\vec{F}\) and \(\vec{G}\) are the flux vectors, and \(\vec{S}\) denotes the chemical source term.

\begin{figure}
    \includegraphics{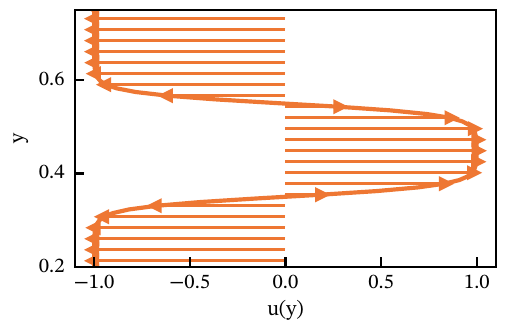}
    \caption{The TDJ velocity profile characterized by two fluid streams moving with counter streamwise velocities.}\label{fig:TDJ_profile}
\end{figure}

The paradigmatic flow considered is  the temporally developing jet (TDJ), shown in \cref{fig:TDJ_profile}.
The TDJ is characterized by a central stream of one species surrounded by another species (e.g., fuel and oxidizer) forming a shear layer with a large  velocity gradient.
The maximum velocity difference across the layers is \(\Delta U = 2U_o\) with \(U_o = \max{(u(x,y,t=0))}\).
The Kelvin-Helmholtz instability~\cite{drazin2002Introduction} in this flow leads to the formation of coherent  vortices,  and subsequent mixing.
Understanding this mixing  is of significant interest in reacting flows in systems such as diesel engines and liquid rocket engines~\cite{mastorakos2009Ignition,martinez-sanchis2023Combustion,wu2010experimental,pitsch2000Scalar,lacaze2012nonpremixed}.

The boundary conditions are assumed to be periodic in both directions, and   the simulation is terminated when the structures reach the top or bottom boundary.
Both reacting and non-reacting flows are considered.
In the former,  \(\Da{} = c_e = 0\), so \(c_1\) and \(c_2\) are conserved scalars.
In the latter,  \(\Da{} > 0\) and the chemical reaction is passive (\(c_e = 0\)) or exothermic (\(c_e > 0\)).
The initial conditions of the velocity components are:
\begin{align}
    u(x,y,0) & = U_o \left( \tanh{\left(\frac{y-y_{\text{min}}}{\delta_i}\right)} - \tanh{\left(\frac{y-y_{\text{max}}}{\delta_i}\right)} - 1\right),                \label{eqn:IC_u} \\
    v(x,y,0) & = 0, \label{eqn:IC_v}
\end{align}
where \(\delta_i\) is a measure of the initial shear layer thickness and \(y_{\text{min}}\) and \(y_{\text{max}}\) denote the locations of the shear layers.
The initial conditions of \(c_1\), \(c_2\), and the temperature \(T\) are
\begin{align}
    c_1(x,y,0) & = \frac{1}{2} \left(\tanh{\left(\frac{y-y_{\text{min}}}{\delta_i}\right)} - \tanh{\left(\frac{y-y_{\text{max}}}{\delta_i}\right)} \right) ,  \label{eqn:IC_c1} \\
    c_2(x,y,0) & = 1 - c_1(x,y,0),\label{eqn:IC_c2}                                                                                                                             \\
    T(x, y, 0) & = 2 \mathcal{A} c_1 (x,y,0).\label{eqn:IC_T}
\end{align}
Here \(\mathcal{A}\) is a constant that specifies how much the temperature increases in the central region  of the flow.
The pressure field is initially constant, which then allows for specification of the initial density field from the equation of state.
The simulation is initiated by adding noise to the shear layers as described in \cref{app:perturbation}.
The criterion for the size of the timestep \(\Delta t \) is described in \cref{app:tstep}.
The numerical solution technique is based on  MacCormack's discretization~\cite{maccormack1969effect} which is second order accurate in both time and space.
\Cref{app:MacCormacks} gives a summary of the method applied to \cref{eqn:gov}.
The simulation of the TDJ is conducted on a \(128 \times 128\) grid with uniform spacing in both the streamwie (\xx{}) and  cross-stream (\yy{}) directions.  With the assumption of a temporally evolving flow, the statistical Reynolds-averaged data are obtained by sampling along the streamwise direction. These are denoted by an overbar. The degree of mixing   is measured by the vorticity thickness \(\delta_\omega = \Delta U / \max{(|\overline{\omega}|)}\), where \(\vec{\omega} = \nabla \times \vec{V}\) denotes the vorticity.
The flow dynamics is also characterized by the Reynolds shear stresses:
$  R = \overline{(u - \overline{u})(v - \overline{v})}$.

\section{MPS Representation and Time Evolution of Transport Variables}\label{sec:MPS}
\subsection{Matrix Product State}\label{sec:MPS_Description}
\begin{figure}
    \includegraphics[scale=1.2]{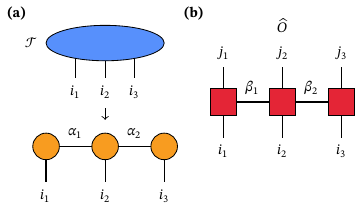}
    \caption{(a) Sketch of an unstructured tensor \(\mathcal{T}\) and the corresponding MPS. Indices \(i_j\) are physical indices. Indices \(\alpha_j\) are a result of the MPS construction and are called bond indices. (b) Three site MPO\@. \(\beta_i\) is used to denote the bond indices to distinguish them from the MPS bond indices.}\label{fig:mps_and_mpo}
\end{figure}

The MPS is a one-dimensional TN in which each tensor is joined to its neighbor, as shown in \cref{fig:mps_and_mpo}(a).
Assume a bitwise representation of the unstructured tensor.
An element of the unstructured tensor is indexed by \(i_1 i_2 \ldots i_N\) where \(N = \log_d(n)\), \(n\) is the number of elements, \(d\) is the physical index size or the size of the local Hilbert space (\(d = 2\) for qubits), and \(i_j = \{0,1,\ldots,d-1\}\) represents the \(j\)th resolved length scale.
For example, with \(n=8\) the number of bits is \(N= 3\), and the fourth element of the tensor would be indexed by \(i_1i_2i_3 = 100\) where \(i_1\) is the least significant bit.
The index \(i\) reflects increasingly fine spatial scales when the tensor represents a  transport  variable.
The first index \(i_1\) is the finest scale, whereas \(i_N\) is the coarsest scale corresponding to splitting the domain into halves.
Each of these spatial scales is isolated as individual tensors in the MPS through a process of successive \(N-1\) SVDs.
A detailed explanation of the construction process is given in Refs.~\cite{schollwoeck2011densitymatrix,catarina2023Densitymatrixa} and specifically for computational fluids in Ref.~\cite{gourianov2022Exploiting}.
After the final SVD, the unstructured tensor becomes
\begin{equation}
    \mT{}_{i_1 i_2 \ldots i_N} =\sum_{\{\alpha_j\}=1}^{\{q_j\}\leq r} M_{\alpha_1}^{i_1} M_{\alpha_1,\alpha_2}^{i_2}\ldots M^{i_N}_{\alpha_{N-1}},\label{eqn:MPS}
\end{equation}
where \(\{\alpha_j\} = \{\alpha_1, \alpha_2, \ldots, \alpha_{N-1}\}\) are the bond indices representing unrealized contractions between tensors, \(q_j = \sizeof{\alpha_j}\) are the sizes of the bond indices, \(r = \max\{q_j\}\) is the maximum bond index size and \(M\) are the tensors of the MPS\@.
The order of a tensor is denoted by the number of legs or indices connected to it.
As shown in \cref{fig:mps_and_mpo}, the interior tensors of the MPS are of the third order and the boundaries are of the second order.
A scalar field in a one-dimensional Cartesian basis, can be expressed as a quantum state of qubits where coefficients are encoded in an MPS.
For example, for the streamwise component of the velocity:
\begin{align}
    \begin{split}
        \ket{u} &= \sum_{\mathclap{i_1,i_2, \ldots , i_N=1}}^d u_{i_1i_2\ldots i_N} \ket{i_1i_2\ldots i_N}\\
        &= \sum_{i_1,i_2,\ldots,i_N}^d\sum_{\{\alpha_j\} = 1}^{\{q_j\}\leq r} M_{\alpha_1}^{i_1} M_{\alpha_1,\alpha_2}^{i_2}\ldots M^{i_N}_{\alpha_{N-1}}\ket{i_1i_2\ldots i_N}.
    \end{split}
\end{align}
Here the \(\ket{\cdot}\) and \(\ket{i_1i_2\ldots i_N}\) are in Dirac notation and correspond to vectors in the Hilbert space~\cite{griffiths2018Introduction}.

\begin{figure}
    \includegraphics{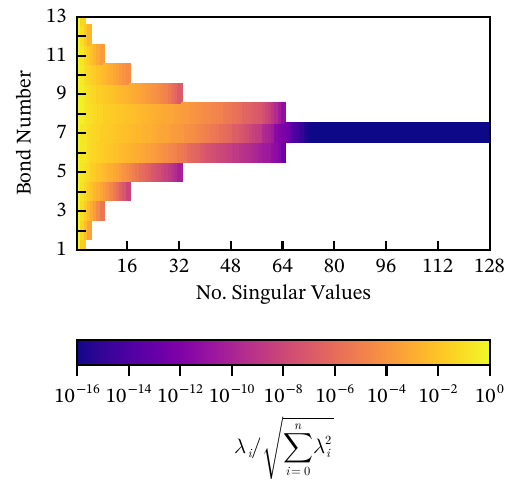}
    \caption{Example of singular value spectrum for the MPS of scalar \(c_1\) with \(\Da{} = 0\), \(\Rey{} = 2500\), and \(\Ma{} = 0.2\). The size of the MPS is \(N = 14\) which corresponds to a \(2^7 \times 2^7 = 128 \times 128\) grid. The colors indicate the relative strength of correlations at each physical scale.}\label{fig:schmidt_example}
\end{figure}

The MPS ansatz readily exposes correlations between the tensors. In the present case, these are correlations between spatial scales of the flow field.
The process of exposing the correlations is done by taking the SVD of the MPS at tensor \(\ell\) when it is in canonical form as described in Ref.~\cite{schollwoeck2011densitymatrix}.
In canonical form, the SVD is equivalent to the Schmidt decomposition~\cite{wilde2016Classical}, and the quantum state is expressed as
\begin{multline}
    \ket{\mT{}} = \sum_{\alpha_\ell}^{q_\ell} S_{\alpha_\ell,\alpha_\ell} \left(\sum_{i_1,\ldots,i_{\ell}} U_{\alpha_\ell}^{i_1,\ldots,i_\ell}\ket{i_1\ldots i_\ell}\right)\\
    \left(\sum_{i_{\ell+1},\ldots,i_{N}} (V^\dagger_{\alpha_\ell})^{i_{\ell+1},\ldots,i_N}\ket{i_{\ell+1}\ldots i_N}\right).\label{eqn:schmdit_decomp}
\end{multline}
In this decomposition, the tensors \(i_1\) to \(i_\ell\) are contracted into a left orthogonal tensor \(U\), and tensors from \(i_{\ell + 1}\) to \(i_N\) are contracted into a right orthogonal tensor \(V^\dagger\).
The singular values \(\lambda_i\) (Schmidt coefficients), which represent the strength of the correlations between these two groupings of tensors, are contained in the diagonal tensor \(S\).
The singular values are organized from largest to smallest along the diagonal of \(S\).
The manner in which \(\lambda_i\) decay can be used to inform an upper bound on the bond index size, denoted by \(\chi\).
Considering all \(\lambda_i\) at every bond constitutes the singular value spectra of the MPS\@.
These spectra directly characterize the correlations among the physical scales of the flow field.
\Cref{fig:schmidt_example} shows the spectrum explicitly for an MPS created from the snapshot of the DNS solution of \(c_1\).
The relative strength of the correlations at a given bond in \cref{fig:schmidt_example} is indicated by the color scale.
Once selected, the bond index \(\chi\) can then be applied to the entire MPS by performing an SVD at each tensor.
The SVD exposes the singular values which are then truncated so that no more than \(\chi\) singular values remain.
The procedure scales as \ord{2dN\chi^3}.
The number of elements in the truncated MPS scales as \ord{dN\chi^2}.
The MPS with bond dimension of \(\chi\) approximates the original unstructured tensor as
\begin{equation}
    \mT{}_{i_1i_2\ldots i_N} \approx \widetilde{\mT{}}_{i_1i_2\ldots i_N} =  \sum_{\{\alpha_j\}=1}^{\{q_j\}\leq \chi}M_{\alpha_1}^{i_1} M_{\alpha_1,\alpha_2}^{i_2}\ldots M^{i_N}_{\alpha_{N-1}}.\label{eqn:MPS_approx}
\end{equation}
The relative truncation error incurred in \cref{eqn:MPS_approx} at the tensor \(i\) is
\begin{equation}
    \varepsilon_i = ||\mT{} - \widetilde{\mathcal{T}}||_2^2 = \frac{\sum_{j=\chi+1}^{\chimax{}} \lambda_{i,j}^2 }{\sum_{j=1}^{\chimax{}} \lambda_{i,j}^2}, \qquad  \varepsilon = \sum_i^{N} \varepsilon_i,\label{eqn:MPS_error}
\end{equation}
where \(\lambda_{i,j}\) is the singular value of \(j\) of the \(i\)th bond and \(\varepsilon\) is the truncation error across all tensors of the MPS\@.
This error (sometimes referred to as the \emph{cutoff}) is distinct from the Taylor Series truncation error in  MacCormack's discretization~\cite{maccormack1969effect}.
This distinction is made clear by referring to the truncation as defined by \cref{eqn:MPS_error} as the \emph{MPS truncation error}.

There are two sets of indices for two-dimensional Cartesian fields, denoted as \(x_1 x_2 \ldots x_N\) and \(y_1 y_2 \ldots y_N\) to contrast with the one-dimensional Cartesian indices of \(i_1 i_2 \ldots i_N\).
There are many possible permutations of the indices within an MPS encoding and the specific choice of scheme is important for MPS truncation error.
In this work the \emph{peak}\footnote{The peak scheme is equivalent to the \emph{comb} scheme described in Refs.~\cite{ye2022Quantuminspired,chepiga2019Comb} in 2D.} scheme is used, in the form
\begin{gather}
    x_1 x_2 \ldots x_N y_N y_{N-1} \ldots y_2 y_1.\label{eqn:index_ordering}
\end{gather}
The peak scheme places the largest scales (\(x_N\), \(y_N\)) in the center of the MPS\@.
This type of ordering was found to minimize the MPS truncation error; therefore  it is used in  all the simulations  (see \cref{app:indexing} for details).
See Refs.~\cite{pisoni2025Compression,tindall2024Compressingb,ye2022Quantuminspired,chepiga2019Comb,gomez-lozada2025Simulating} for additional permutations.

\subsection{Characterization of MPS Truncation}
The MPS representation is not unique, which allows canonical forms which trivialize otherwise costly operations~\cite{schollwoeck2011densitymatrix}.
However, non-uniqueness also means that there is an overhead in the number of parameters of the MPS, \(P_T\):
\begin{equation}
    P_T = d\sum_1^N \sizeof{\alpha_{i-1}}\sizeof{\alpha_i}. \label{eqn:total_parameters}
\end{equation}
Thus, \(N=3\) and \(d = 2\) give \(P_T = 16\) which is double the number of elements of the Cartesian representation.
\(P_T\) scales as \ord{d^{N/2}}.
\Cref{eqn:total_parameters} can be used to define a threshold above which MPS is not competitive with DNS~\cite{ye2022Quantuminspired}.
The value of \(\chi\) which gives an MPS with the same number of parameters as the DNS is denoted as \chicompete{}.
The number of DNS parameters is equal to \(n_x n_y\) where \(n_x\) and \(n_y\) denote the number of grid points in each coordinate direction.
\Cref{eqn:total_parameters} equals \(\sim{}128^2\) when \(\chi = 44\) so that \(\chicompete{} = 44\) for \(N = 14\).
Based on \cref{eqn:total_parameters} \chicompete{} scales as \ord{N^{2/5}}.

The degree to which the MPS is able to truncate the DNS solution is measured by the truncation ratio as
\begin{equation}
    K = \frac{P_T(\widetilde{\mT{}})}{n_x n_y},\label{eqn:compression}
\end{equation}
where the lower the value of \(K\), the larger the truncation.
An MPS with no truncation (\(\chi = \chimax{}\)) results in \(K > 1\).
When \(\chi = \chicompete{}\) then \(K\approx 1\).
Another measure of truncation is based on the number of degrees of freedom (dof) of the MPS defined as~\cite{holtz2012manifoldsa}
\begin{equation}
    \dof{} = P_T - P_G, \qquad P_G = \sum_{i=1}^{N-1} \alpha_i^2,\label{eqn:dof}
\end{equation}
where \(P_G\) denotes the gauge dof.
In DNS, the total parameters are equal to the \(\dof{} = n_x n_y = 128^2\).
For an MPS with \(\chicompete{} = 44\) the dof is 7568 which is approximately 46\% of the DNS.
The truncation based on \cref{eqn:dof} (in contrast to \cref{eqn:total_parameters}) is considered the theoretical best case truncation.
In practice, truncation ratios based on \dof{} may be unrealized because of the accumulation of MPS truncation error (see \cref{sec:compression_error} for details).

The error incurred by a truncated MPS is given by \cref{eqn:MPS_error} if the singular values between the original and truncated forms are the same up to \(\chi\).
In general, the singular values will not be the same for MPS that have been time evolved because of the accumulation of MPS truncation error.
Therefore, another metric is needed to estimate the error.
The fidelity \(\mathcal{F}\) measures how similar two MPS are and is equal to ~\cite{wilde2016Classical}
\begin{equation}
    \mathcal{F}\left(\psiMPS{}, \psiDNS{}\right) = \frac{\left|\braket{\psiMPS{}|\psiDNS{}}\right|^2}{\braket{\psiMPS{}|\psiMPS{}}\braket{\psiDNS{}|\psiDNS{}}}.\label{eqn:fidelity}
\end{equation}
Here \psiMPS{} is the time evolved truncated MPS, \psiDNS{} is formulated directly from the DNS solution with no truncation (i.e., \(\chi = \chimax{}\)), and \(\braket{\cdot|\cdot}\) denotes the inner product between two states in Dirac notation.
Perfect agreement corresponds to \(\mathcal{F} = 1\).
The error between the MPS and DNS solutions can then be quantified by the infidelity defined as  \(\mathcal{I} = 1 - \mathcal{F}\).
If both MPS are formed directly from the DNS solution so that the truncated MPS is \(\ket{\psiMPS{}} =\ket{\widetilde{\psiDNS{}}}\) then \cref{eqn:fidelity} reduces to
\begin{equation}
    \mathcal{F}\left(\ket{\widetilde{\psiDNS{}}},\ket{\psiDNS{}}\right) = \frac{\sum_{i}^\chi  \lambda_{\text{DNS},i}^2 }{\sum_i^{\chimax{}} \lambda_{\text{DNS},i}^2},\label{eqn:fidelity_reduced}
\end{equation}
and \(\mathcal{I}\) is equivalent to \cref{eqn:MPS_error}.
Generally this is not the case because of errors, which manifest as differences in the singular values \(\lambda_i\), introduced by time evolution of the truncated MPS\@.
\subsection{Matrix Product Operators}\label{sec:MPO}
The matrix product operator (MPO) is the MPS analogue of an operator matrix~\cite{schollwoeck2011densitymatrix} for representing spatial derivatives.
These derivatives are evaluated via finite difference approximation and are arranged into a sparse operator matrix form.
The corresponding MPO is constructed in the same manner as described for the MPS\@.
The number of physical indices per tensor is doubled because the MPO has an input and output (corresponding to the rows and columns of an operator matrix).
A general MPO has the form \cite{schollwoeck2011densitymatrix}:
\begin{equation}
    \MPO{} = \sum_{\mathclap{\substack{j_1,j_2,\ldots,j_N\\i_1,i_2,\ldots,i_N\\\beta_1,\beta_2,\ldots,\beta_{N-1}}}} \mpo{\beta_1}{j_1,i_1} \mpo{\beta_1,\beta_2}{j_2,i_2}\ldots \mpo{\beta_{N-1}}{j_N,i_N} \ket{j_1j_2\ldots j_N}\bra{i_1i_2\ldots i_N},\label{eqn:MPO}
\end{equation}
as shown diagrammatically for \(N=3\) in \cref{fig:mps_and_mpo}(b).
The maximum bond index size of the MPO is denoted by \(C\); generally \(C \ll \chi\).
The MPO \(\MPO{}\) is applied to an MPS \mpsA{} to generate a new MPS \mpsB{}:
\begin{gather}
    \mpsB{} = \MPO{}\mpsA{} = \sum_{\mathclap{\substack{j_1,j_2,\ldots,j_N\\\gamma_1,\gamma_2,\ldots,\gamma_{N-1}}}} \res{\gamma_1}{j_1} \res{\gamma_1,\gamma_2}{j_2} \ldots\res{\gamma_{N-1}}{j_N} \ket{j_1j_2\ldots j_N},\label{eqn:mpo-mps}
\end{gather}
where
\begin{equation}
    B_{\gamma_{i-1},\gamma_i}^{j_i} = A^{i_i}_{\alpha_{i-1},\alpha_{i}} O_{\beta_{i-1},\beta_i}^{i_i,j_i}.
\end{equation}
Here the bond indices of \(\MPO{}\) and \mpsA{} are combined to form a new bond index for \mpsB{} defined as \(\gamma_i = \alpha_i \beta_i\).
The growth of the bond index size as \(\chi \rightarrow \chi C\) requires the MPS be regularly truncated after MPO-MPS operations in order to remain tractable.
This is accomplished using the same process of SVD sweeps described in \cref{sec:MPS_Description}.
The cost of the MPO-MPS product scales as \ord{Nd^2\chi^2C^2} assuming a tensor-by-tensor contraction.
The aforementioned SVD sweeps scale as \ord{N\chi^3} so that the overall cost of the MPO-MPS product is \ord{N\chi^3}.
\subsection{MPS Algorithm}\label{sec:MPS_Operations}
The MPS algorithm for time evolution follows the finite difference implementation described in \cref{app:MacCormacks}, known as the MacCormack's method. Here, the variables and operators are replaced by the MPS/MPO equivalents as shown in \cref{fig:alg} for the \(U_2\) component of \cref{eqn:gov}, corresponding to the \xx{} component of velocity.
A constant \(\chi\) is applied to every MPS throughout the entire simulation.
Importantly, \(\chi\) is also enforced for all intermediate quantities (e.g., \(\tau_{xx}\)).
Ideally, \(\chi\) would be allowed to dynamically adjust as needed based on the correlations in the simulation.
In practice this is challenging for a system of many variables (\(\rho\), \(u\), \(v\), etc.) at small sizes (\(N = 14\)) and with many operations per time step.
The combination of these factors leads to unacceptably large rates of MPS truncation error accumulation.

All required mathematical operations needed to mirror the finite difference implementation are possible in the MPS ansatz.
The costs associated with each are summarized in \cref{tab:costs}.
MPO-MPS operations are done using the algorithm described in Ref.~\cite{verstraete2004Renormalization} (referred to as the \textit{fit} algorithm in~\cite{stoudenmire2010Minimally}, see also Refs.~\cite{mcculloch2007densitymatrixa,schollwoeck2011densitymatrix,camano2025Successive} for additional methods).
Finite difference MPOs with \(C = 2 \text{ and } 3\) are constructed using the process described in \cref{app:MPO_construction}.
Two important operations, Hadamard multiplication and division, are described in \cref{app:hadamard_mult,app:hadamard_div}, respectively.
\begin{algorithm}[t]
    \KwData{Encode fields as MPS}
    \(\ket{\rho} \gets \rho \quad \ket{u} \gets u \quad \ket{v} \gets v \quad \ket{p} \gets p\)\;
    \(\ket{T} \gets T \quad \ket{c_1} \gets c_1 \quad \ket{c_2} \gets c_2\)\;
    \KwData{Construct MPOs}
    \(\dx{}\gets \partial / \partial x \quad \dy{} \gets \partial / \partial y\)\;
    \Begin{%
        \(\ket{U_2} = \ket{\rho} \odot \ket{u}\)\;
        \For{\(i \in tsteps\)}{%
            Compute predictor flux vectors\;
            \(\ket{F_2} = \ket{U_2} \odot \ket{u} \oplus \ket{p} \ominus \frac{1}{\Rey{}}\ket{\tau_{xx}}\)\;
            \(\ket{G_2} = \ket{U_2}\odot\ket{v} \ominus \frac{1}{\Rey{}}\ket{\tau_{yx}}\)\;
            Compute predictor solution vector\;
            \(\ket{\pfrac{U_2}{t}} = - \dx{} \ket{F_2} \ominus \dy{} \ket{G_2} \oplus \ket{R} \)\;
            \(\ket{\overline{U}_2} = \ket{U_2} \oplus \Delta t \ket{\pfrac{U_2}{t}} \)\;
            Update primitive variables\;
            \(\ket{\overline{u}} = \ket{\overline{U}_2}\oslash \ket{\overline{U}_1}\)\;
            Compute corrector flux vectors\;
            \(\ket{\overline{F}_2} = \ket{\overline{U}_2} \odot \ket{\overline{u}} \oplus \ket{\overline{p}} \ominus \frac{1}{\Rey{}}\ket{\overline{\tau}_{xx}}\)\;
            \(\ket{\overline{G}_2} = \ket{\overline{U}_2}\odot\ket{\overline{v}} \ominus \frac{1}{\Rey{}}\ket{\overline{\tau}_{yx}}\)\;
            Compute corrector solution vector\;
            \(\ket{\pfrac{\overline{U}_2}{t}} = - \dx{} \ket{\overline{F}_2} \ominus \dy{} \ket{\overline{G}_2} \oplus \ket{\overline{R}} \)\;
            \(\ket{\pfrac{U_2}{t}}_{av} = \frac{1}{2}\left( \ket{\pfrac{U_2}{t}} \oplus \ket{\pfrac{\overline{U}_2}{t}} \right)\)\;
            \(\ket{U_2}^{n+1} = \ket{U_2} \oplus \Delta t \ket{\pfrac{U_2}{t}}_{av}\)\;
            Update primitive variables\;
            \(\ket{u}^{n+1} = \ket{U_2}^{n+1}\oslash \ket{U_1}^{n+1}\)\;
        }
    }
    \caption{MPS Algorithm for MacCormack's Method for the \(U_2\) Component of the Solution Vector.}\label{fig:alg}
\end{algorithm}
\begin{table}[b]
    \caption{Summary of operations between \mpsA{} and \mpsB{}.}\label{tab:costs}
    \begin{tabular}{l l l}
        \toprule
        Operation                                                         & Cost         & Bond Growth                \\
        \midrule
        Addition (Direct Sum \cite{stoudenmire2010Minimally})             & \ord{\chi^3} & \(\chi = \chi_A + \chi_B\) \\
        Differentiation (MPO-MPS via fit \cite{stoudenmire2010Minimally}) & \ord{\chi^3} & \(\chi = \chi_A C \)       \\
        Multiplication (see \cref{app:hadamard_mult})                     & \ord{\chi^4} & \(\chi = \chi_A \chi_B\)   \\
        Division (see \cref{app:hadamard_div})                            & \ord{\chi^4} & \(\chi = \chi_A \chi_B\)   \\
        \bottomrule
    \end{tabular}
\end{table}
\section{Simulation of Compressible and Reacting Flows}\label{sec:results}
Simulations are conducted of both non-reacting and reacting conditions.
In the former the MPS algorithm is validated and the effects of compressibility are assessed.
In the latter, the influence of reaction exothermicity on fluid dynamics is investigated.
\subsection{Validation}
\begin{figure}[b]
    \includegraphics{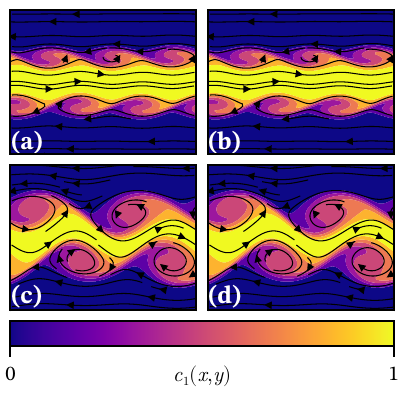}
    \caption{MPS algorithm validation case with \(\chi = \chimax{} = 128\) showing contours of \(c_1\). The black arrows show the velocity field \(\vec{V}\). (a) and (c) are the DNS solution at \(t = 0.8\) and \(t = 1.2\), respectively. (b) and (d) are the MPS solution  at \(t = 0.8\) and \(t = 1.2\), respectively.}\label{fig:valid_contours}
\end{figure}

The MPS algorithm is validated by performing simulations without truncations (\(\chi = 128\)) and comparing
the results with those obtained via DNS.
These simulations are conducted with \(\Rey{} = \Pe{} = 2500\), \(\Ma{} = 0.2\), and \(\Da{} = 0\).
% The Lewis Number was taken as identity, so no distinction is made in the \Pe{} with respect to thermal and mass diffusion.
\Cref{fig:valid_contours} shows the contours of the conserved scalar \(c_1\) at two different times and overlaid with the velocity field \(\vec{V}\).
% The infidelity and vorticity thickness provide a more quantitative measure of agreement as shown in \cref{fig:valid_metrics}.
An excellent agreement is observed, with the infidelity of all primitive variables not exceeding \num{e-14} after \(\sim{}\)2260 timesteps.
\subsection{Impact of Compressibility}
\begin{figure}
    \includegraphics{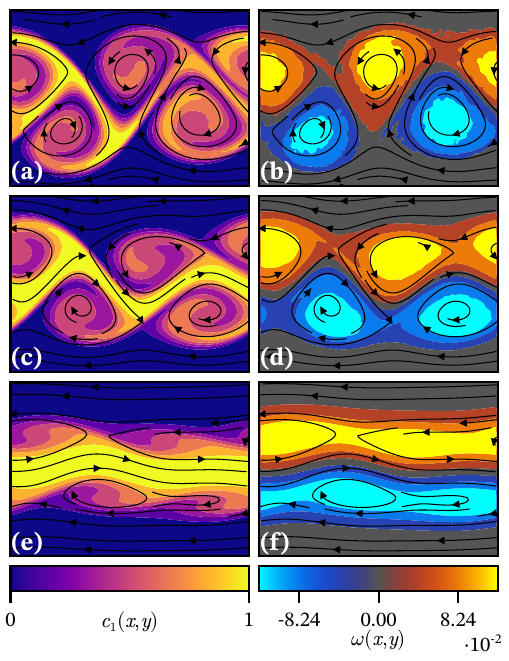}
    \caption{Non-reacting (\(\Da{} \!=\! 0\)) MPS simulation with  \(\chi = 34\) at \(t \sim{} 2.7\). (a), (c), and (e) are \(c_1\) contours at \(\Ma{} =\numlist{0.2;0.4;0.6}\), respectively. (b), (d), and (f) are \(\omega\) at \(\Ma{} =\numlist{0.2;0.4;0.6}\), respectively. The black arrows show the velocity field \(\vec{V}\).}\label{fig:compress_contours}
\end{figure}

\begin{figure}
    \includegraphics{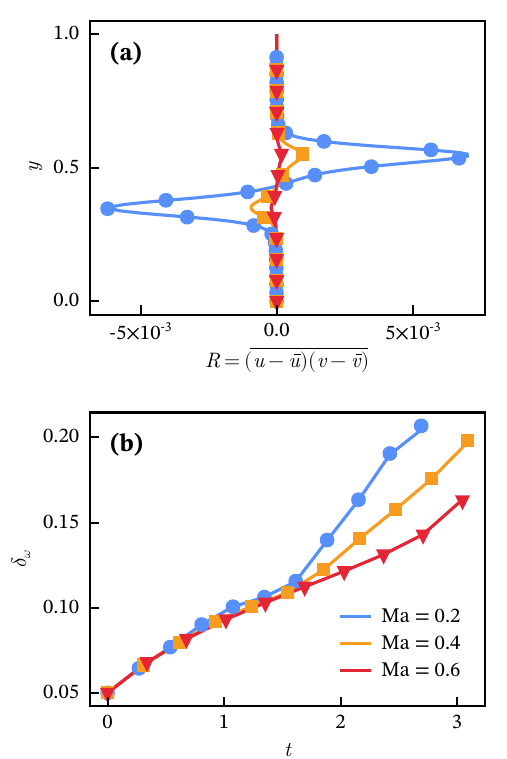}
    \caption{The impact of compressibility as characterized by the Reynolds stresses (a) and the vorticity thickness (b) for non-reacting MPS simulations with \(\chi = 34\). The shear stresses are shown for \(t = 1.4\) (\(\Ma{} = 0.2\)) and \(t= 1.5\) (\(\Ma{} = 0.4\) and \(\Ma{} = 0.6\)). DNS results are identified via symbols overlaid on the MPS solution (lines).}\label{fig:compress_metrics}
\end{figure}

A set of non-reacting TDJ simulations are performed at \(\chi = 34\) (\(K \sim{} 0.71\)), \(\Rey{} = \Pe{} = 2500\), and with varying Mach number.
The results are summarized in  \cref{fig:compress_contours,fig:compress_metrics}.
In \cref{fig:compress_contours} the left column shows the contours of the conserved scalar \(c_1\) for different values of \Ma{}.
The right column shows the contours of the vorticity.
As \Ma{} increases, the development and growth of the vortical structure is suppressed, with almost none observed for  \(\Ma{} = 0.6\).
The temporal evolution of Reynolds stresses and the vorticity thickness is shown in \cref{fig:compress_metrics} and reflect excellent agreement with DNS\@.
The magnitude of Reynolds stress increases (stronger turbulent fluctuations, enhanced mixing) as \Ma{} decreases.
The same trend is observed for vorticity thickness wherein a lower \Ma{} results in a faster growth of the layer.

\begin{figure}
    \includegraphics{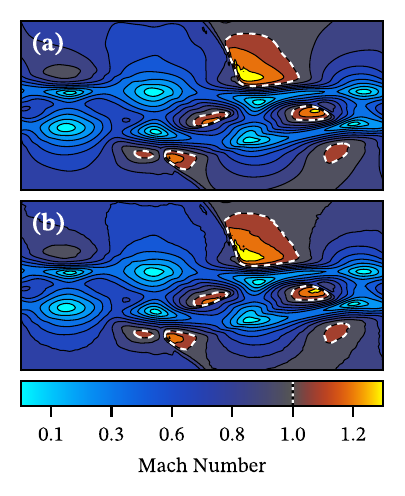}
    \caption{Development of shocklets at \(t = 1.8\) in the Mach field \(\Maa{}(x,y)\) for (a) DNS and (b) MPS. The MPS simulations are  conducted with \(\Ma{} = 0.8\), \(Re{} = 5000\), \(N=14\), and with \(\chi = 54\) corresponding to \(K\sim{}1.3\) and \(\dof{} =  9828\). The infidelity of the MPS simulation is \(\sim{}\num{e-4}\).}\label{fig:shocklets}
\end{figure}

Shock waves are another important phenomenon in compressible flows, particularly in high speed transport, explosions, and ballistics.
Shocks form in regions of the flow with large gradients in pressure, density, temperature, and velocity.
Correspondingly, the prediction of shocks informs various aspects of design, for example in aircraft shocks affect aeroacoustics, flow separation, and drag.
A manifestation of shocks are shocklets that are essentially small localized shock waves~\cite{freund2000Compressibility}.
\Cref{fig:shocklets} demonstrates that the MPS is capable of reproducing this important phenomenon.
The instantaneous Mach field shown in \cref{fig:shocklets} is defined as \(\Maa{}(x,y) = |\vec{V}(x,y)| / a(x,y)\) where \(a(x,y)\) denotes the local speed of sound.
It is usually challenging to capture shocks in CFD because of the formation of steep gradients in the transport variables.
Here, in the MPS anstaz with \(\chi\) =  54, it is possible to capture this complex feature of the flow with only ~60\% of the dof of DNS\@.
\subsection{Exothermicity Effects}\label{sec:heat_release}
\begin{figure}
    \includegraphics{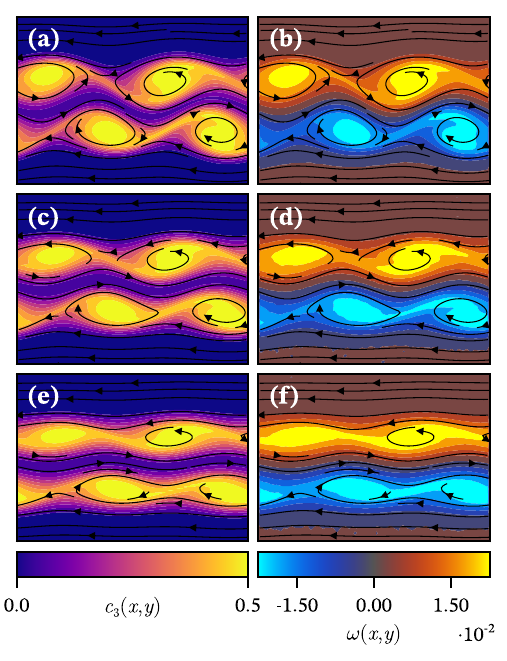}
    \caption{Results for reacting simulation with MPS of \(\chi = 34\), \(\Ma{} = 0.2\), \(\Rey{}= \Pe{} = 2500\), and \(t\sim{}1.9\). (a), (c), and (e) are \(c_3\) contours at \(c_e = \numlist{0;0.3;0.6}\), respectively. (b), (d), and (c) are \(\omega\) at \(c_e = \numlist{0;0.3;0.6}\), respectively. The black arrows show the velocity field \(\vec{V}\).}\label{fig:heat_releases_contours}
\end{figure}

\begin{figure}
    \includegraphics{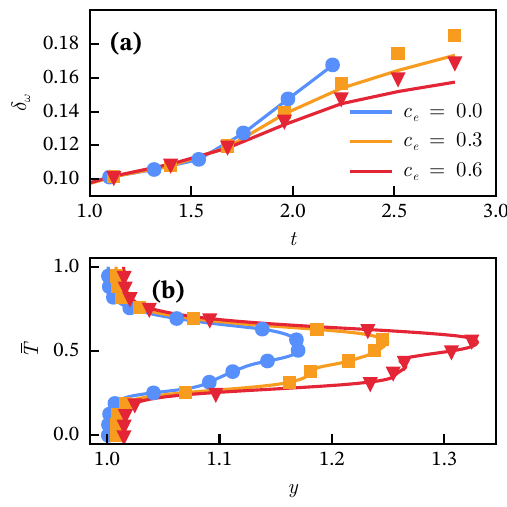}
    \caption{Impact of heat release as characterized by (a) vorticity thickness and (b) streamwise averaged temperature at \(t = 2.7\) for reacting MPS simulations with \(\chi = 34\). The solution for \(c_e =0.3\) and \(c_e =0.6\) deviates from the DNS after \(t\approx 2.0\). This is a manifestation of the accumulation of MPS truncation error. DNS results are shown as symbols overlaid on the MPS solution (lines).}\label{fig:heat_releases_metrics}
\end{figure}

The impact of heat release is captured by keeping the \Damkohler{} number fixed at \(\Da = 1\) and varying the heat release parameter \(c_e\).
The added energy increases temperature and decreases the mean density, which then slows the growth of the shear layer and reduces mixing~\cite{day1998structure,givi1991Effects,shin1991Effect}.
Thus, similar to the impact of compressibility, the exothermicity acts counter to mixing of the two fluid streams of the TDJ\@.
\Cref{fig:heat_releases_contours} illustrates this phenomenon.
As \(c_e\) increases, the degree to which vortices develop and grow decreases, implying a decreased mixing of the reactants.
The delay in mixing is reflected in the MPS as shown in the vorticity thickness in \cref{fig:heat_releases_metrics}(a).
After \(t\gtrsim 2\) the MPS is observed to deviate from DNS for \(c_e \geq 0.3\) due to the accumulation of MPS truncation error.
The error in the heat release flow simulations is expected to grow faster than the error in the non-reacting flow.
The reason is that the effect of heat release is to couple chemistry with hydrodynamics.
Small errors in \(c_1\) and \(c_2\) propagate to \cref{eqn:cons_of_momentum,eqn:cons_of_energy} via the internal energy \(e\).
\Cref{fig:heat_releases_metrics}(b) shows that the highest average temperature occurs for \(c_e = 0.6\).
The validity of the MPS results is verified using an energy balance of the form \(T \approx  1 + c_3 c_e\) at low \Ma{}.
%Here it has been assumed the kinetic energy is much less than the specific internal energy (which is a valid assumption for low \Ma{} flows) and the total energy added from heat release is small.
The relative error in \(T\) with respect to DNS does not exceed \(0.4\%\) in all simulations.
\subsection{MPS Truncation Error}\label{sec:compression_error}
The MPS truncation error, given by \cref{eqn:MPS_error}, grows with time and leads to increasing disagreement between the MPS simulation and the DNS\@.
Eventually, the MPS simulation crashes at which point the system is said to have reached the ``runaway time''~\cite{gobert2005Realtimea}.
The MPS truncation error can be reduced by increasing \(\Delta t\) (decreasing the number of time steps), decreasing the number of operations per time step, and increasing \(\chi\).
Increasing \(\Delta t\) trades the MPS truncation error for the finite difference truncation error (MacCormack's is \ord{\Delta t^2}) and is constrained by the Courant–Friedrichs–Lewy condition~\cite{hoffmann2004Computational}.
The number of operations per timestep can be reduced by selecting other discretization schemes.
Increasing \(\chi\) reduces the error by better capturing the strong correlations that spread across many tensors.
These correlations correspond to slow decaying singular values which increase the magnitude of the summation in the numerator of \cref{eqn:MPS_error}.
However, increasing \(\chi\) beyond \chicompete{} acts counter to the goal of TNs which is to reduce the  computational memory.
It is possible that the \(\chi\) required to achieve a given infidelity threshold reaches an upper bound even as the system size continues to grow.
For example, such an upper bound is demonstrated for the density and velocity fields in \cref{sec:mps_advantage}.
In such a scenario, the correlations in the fields have a behavior akin to the area law of many-body quantum systems~\cite{verstraete2008Matrixa}, and the MPS truncation error from \(\chi\) becomes smaller as \(N\) increases.

\begin{figure}
    \includegraphics{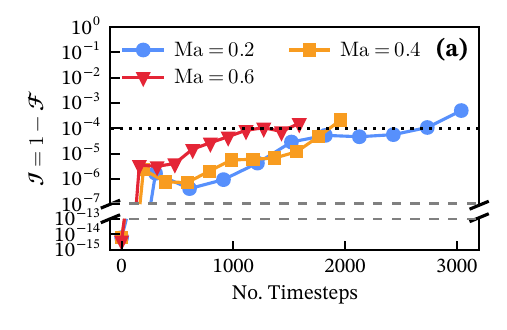}\vspace{-5mm}
    \includegraphics{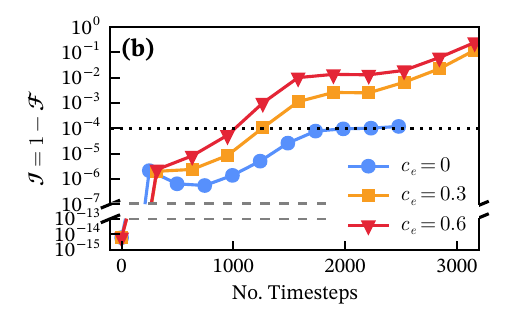}
    \caption{
        MPS truncation error as estimated by infidelity \(\mathcal{I}\) for simulations with \(\chi = 34\).
        The maximum infidelity is shown considering all the primitive variables at each timestep for (a) non-reacting (\(\Da{} = 0\)) and (b) reacting (\(\Da{} > 0\)) simulations.
        A threshold of \num{e-4} is indicated by the dotted line.
        The non-reacting simulations in (a) remain at or below this threshold for all timesteps and reflect excellent agreement with DNS\@.
        For reacting simulations in (b) those with heat release (\(c_e > 0\)) exceed this threshold for some primitive variable fields.
        Reacting flow simulations with no heat release (\(c_e = 0\)) remain at or below the threshold.
        At \(t = 0\) the primitive variable fields are either constant or trigonometric functions that can be represented exactly with \(\chi \leq 10\).}\label{fig:compression_error}
\end{figure}

The combination of these factors (timestep size, operations per timestep, and \(\chi\)) can limit the MPS truncation ratio.
Here,  the selection of \(\chi\) and timestep size is determined iteratively via numerical experimentation.
Typically, the size of \(\Delta t\) is made as large as possible without violating the CFL limit while simultaneously decreasing \(\chi\).
For simulations without heat release (\(c_e = 0 \)) at \(\chi = 34\) the worst case (largest) infidelity of all primitive variables remains at \(\sim{}\num{e-4}\) as shown in \cref{fig:compression_error}.
For simulations with heat release (\(c_e > 0\)), the infidelity of the primitive variables exceed \num{e-4} and lead to small deviations in the solution after approximately 2200 timesteps or \(t\sim{}2\), as shown in \cref{fig:compression_error,fig:heat_releases_metrics}.

The impact of reducing the operations per timestep can be studied by applying the MacCormack's method to a smaller system of PDEs.
The algorithm described in \cref{sec:MPS_Operations} requires 52 MPO-MPS products, 88 direct sums, 37 Hadamard products, and 10 Hadamard divisions per timestep.
Here, the calculations are repeated, but only for \(c_1\) and \(c_2\) which reduces the number of operations per timestep to 16 MPO-MPS products, 24 direct sums, 10 Hadamard products, and 4 Hadamard divisions.
The velocity and density fields are known \textit{a priori}.
The lower number of operations per time step decreases the rate of MPS' error accumulation and makes it possible to control \(\chi\) with cutoff (see \cref{eqn:MPS_error}) instead of a constant value.
Thus, \(\chi\) varies dynamically in this simulation based on a cutoff of \num{e-7}.
The total number of parameters of \(c_1\) or \(c_2\) do not exceed \(\sim{}5600\) throughout the simulation and \(\chi \leq 20\).
The compression ratio for all times is approximately \(K \sim{} 0.31\).
The infidelity for \(c_1\) and \(c_2\) is \(\lesssim \num{e-4}\) at all times.
\begin{figure}
    \includegraphics{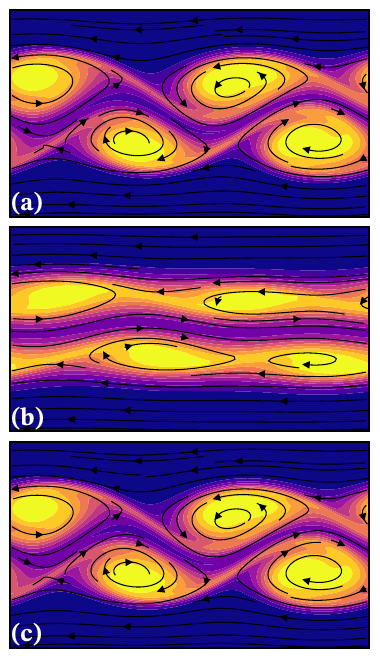}
    \caption{Compressible passive reacting flow simulations with \(\Da{} = 1\), \(\Rey{} = \Pe{} = 2500\), \(c_e = 0\), and \(\Ma{} = 0.2\) via  (a)  DNS, (b)  URDNS (b), and  (c) MPS\@. The MPS simulations are  performed with \(\chi{} = 34\) which has 5508 dof according to \cref{eqn:dof} and a compression of \(K\sim{}0.7\) (based on \(P_T\)). The equivalent URDNS grid is \(74 \times 74\). The DNS grid size is \(128 \times 128\).  The URDNS simulation fails to accurately describe the influence of mixing on the reactant conversion rate.}\label{fig:high_damkohler_contours}
\end{figure}
\begin{figure}
    \includegraphics{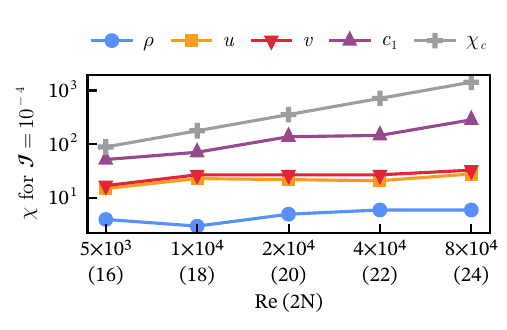}
    \caption{The \(\chi\) required to achieve an infidelity of \num{e-4} based on \textit{a posteriori} analysis of DNS\@. The MPS size is shown in parentheses in the \xx{}-axis (as in \cref{fig:dof_vs_PT}). \chicompete{} and \(c_1\) scale as \ord{\Rey^{0.017}}  and \ord{\Rey^{0.002}}, respectively. \(\rho\), \(u\), and \(v\) remain approximately constant with increasing \Rey{}.}\label{fig:reynolds_scaling}
\end{figure}
\begin{figure}
    \includegraphics{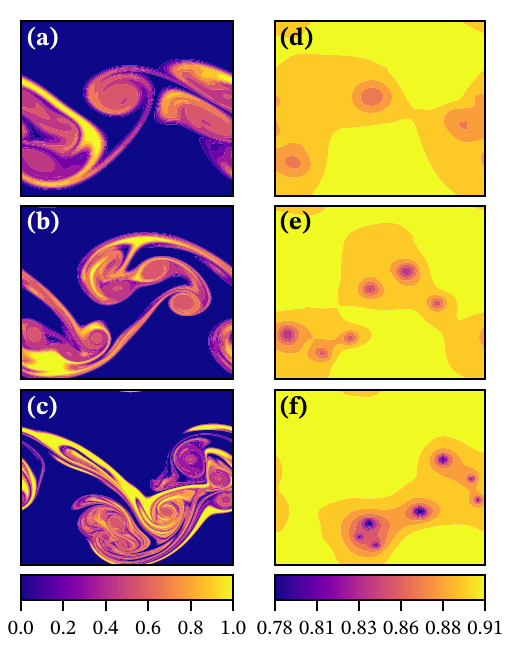}
    \caption{Contours of \(c_1\) (a, b, c) and \(\rho\) (d, e, f) at \(t = 1.6\) from a DNS for \(\Ma{} = 0.2\) and \(\Da{} = 0\). (a,d) \(\Rey{} = \num{5e3}\) on a \(256\times 256\) grid, (b,e) \(\Rey{} = \num{20e3}\) on a \(1024\times 1024\) grid, (c,f) \(\Rey{} = \num{80e3}\) on a \(4096\times 4096\) grid. Increasingly fine structures manifest in \(c_1\) as the Reynolds number grows.}\label{fig:reynolds_scaling_contours}
\end{figure}
\subsection{Advantage of MPS}\label{sec:mps_advantage}
A means of demonstrating the  advantage of the MPS ansatz is to measure its  ability to evolve a flow field with less dof than the DNS\@.
Consider the passive reaction (\(c_e = 0.0\) and \(\Da{} = 1\)) from \cref{sec:heat_release}.
With \(\chi = 34\) the dof according to \cref{eqn:dof} is \num{5508}.
An underresolved DNS (URDNS) with approximately the same dof is given by \(n' = \sqrt{\dof{}} \approx 74\) where \(n'\) is the grid size.
The solutions via  DNS, URDNS, and MPS are compared in \cref{fig:high_damkohler_contours} at \(t = 3.2\).
As observed, the URDNS does not capture the correct dynamics.

\begin{figure}
    \includegraphics{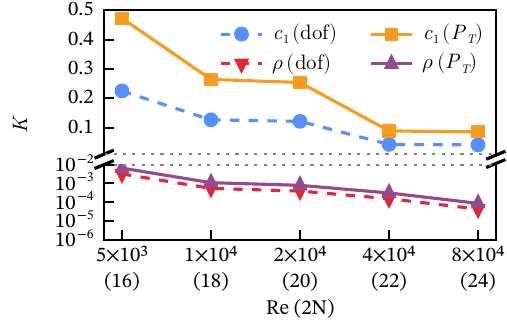}
    \caption{Best (\(\rho\)) and worst (\(c_1\)) truncation ratios \(K\) for \(P_T\) and dof based on analysis of DNS\@. The MPS size is shown on the \xx{}-axis in parentheses. At each snapshot of the DNS, the bond dimension \(\chi\) as predicted by \cref{fig:reynolds_scaling} is  applied. The truncation of \(c_1\) based on total parameters \(P_T\) approaches that of the dof as \(N\) increases. For \(\rho\) the truncation based on \(P_T\) has converged to that of dof.}\label{fig:dof_vs_PT}
\end{figure}

\Cref{fig:high_damkohler_contours} demonstrates that the MPS is using the available dof much more efficiently.
However, the small system size of \(N=14\) limits the MPS truncation so that the \emph{number of parameters} is \num{11701} which corresponds to \(K\sim{}0.70\).
The challenge then becomes one of decreasing the number of parameters so that \(P_T \rightarrow \dof{}\) and thus \(K \rightarrow 0\).
This becomes increasingly achievable as \(N\) increases.
To emphasize this point, a series of DNS are performed at larger Reynolds numbers and grid sizes, corresponding to increasing \(N\).
DNS solutions are extracted and expressed in the MPS ansatz.
For those MPS, the \(\chi\) needed to achieve an infidelity of \num{e-4} is shown in \cref{fig:reynolds_scaling} as a function of Reynolds number and \(N\).
The \(\chi\) required for each variable shown grows much slower than \chicompete{}.
Specifically, \(\chi\) for \(\rho\), \(u\), and \(v\) stay approximately the same, while for \(c_1\) it increases with Reynolds number as \ord{\Rey^{0.002}}.
This increase is a consequence of the complexity in the scalar field as the Reynolds number increases.
Such complexity is illustrated in \cref{fig:reynolds_scaling_contours} and is absent in \(\rho\), \(u\) and \(v\).
In particular, the \(\chi\) for \(c_1\) grows slower than \chicompete{}.
The best and worst truncation ratios, corresponding to \(\rho\) and  \(c_1\), respectively, are shown in \cref{fig:dof_vs_PT} and confirm that \(P_T\) can approach \dof{}.
It should be noted that the MPS truncation error is absent in \cref{fig:reynolds_scaling,fig:dof_vs_PT} because the MPS in these figures is formed directly from the DNS solution.
However, were the MPS truncation error present it would decrease as the system size \(N\) increases because \(\chi\) grows only moderately in the worst case (\(c_1\)) as shown in \cref{fig:reynolds_scaling}.
This means that increasingly more of the truncated singular values are smaller in magnitude (i.e., the error is decreasing).

\begin{figure}
    \includegraphics{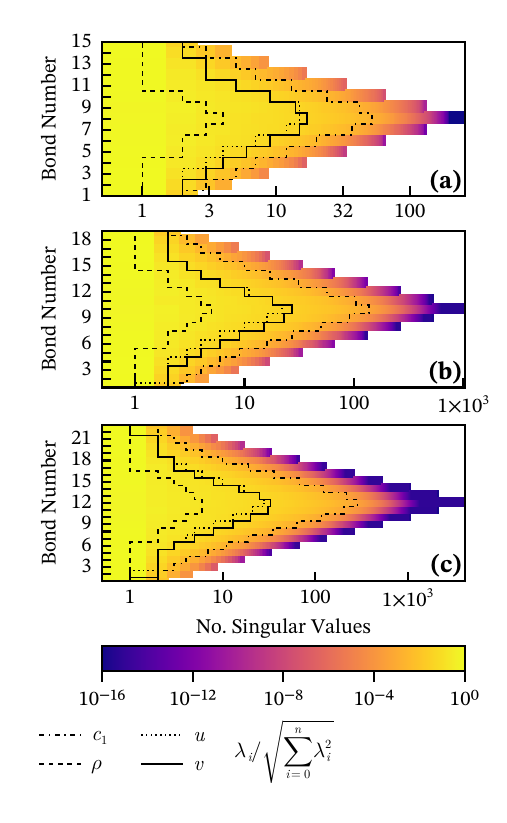}
    \caption{Normalized singular value spectra of \(c_1\), \(\rho\), \(u\), and \(v\) based on DNS snapshots at \(t=1.6\) shown in \cref{fig:reynolds_scaling_contours} where (a) \(\Rey{} = \num{5e3}\), (b) \(\Rey{} = \num{20e3}\), and (c)  \(\Rey{} = \num{80e3}\). The MPS are constructed so that the infidelity is \(\mathcal{I} \leq \num{e-4}\). The DNS (full spectrum, no truncation) is shown in color for \(c_1\) and has a pyramid shape from the SVD construction  (e.g., see \cref{fig:mps_and_mpo}). The bond dimensions of the MPS for \(c_1\), \(u\), \(v\), and \(\rho\) are shown as black lines. The required \(\chi\) stays relatively constant for \(\rho\), \(u\) and \(v\) with increasing Reynolds number (increasing MPS size \(N\)) but increases moderately for \(c_1\).}\label{fig:reynolds_schmidt}
\end{figure}

Other fields such as density and velocity achieve significantly larger truncation compared to \(c_1\) as shown in the singular value spectra in \cref{fig:reynolds_schmidt}.
For example, for \(\rho\), \(u\), and \(v\), \(\chi \leq 44\) for \(\Rey{} = 8 \times 10^{4}\) (\(r = 4096\)) which corresponds to a truncation ratio of 1:300 (based on total parameters).
Another means of  minimizing \(P_T\) is to dynamically select \(\chi\) on a per-bond basis, as shown in \cref{fig:reynolds_schmidt}, rather than applying a constant \(\chi\) as done here.
\Cref{fig:reynolds_schmidt} illustrates the impact of the Reynolds number on the Schmidt spectrum.
The increasing complexity of \(c_1\) with \Rey{}, evidenced in \cref{fig:reynolds_scaling_contours}, is reflected in an increasingly large \(\chi\) in \cref{fig:reynolds_schmidt}.
The dynamic selection of \(\chi\) has the advantage of growing or shrinking \(\chi\) as dictated by the way correlations between scales change as time evolves.
This selection can be performed by applying a cutoff at each bond as defined by \cref{eqn:MPS_error} and as  done for the passive scalars (\(c_e = 0 \)) as described in \cref{sec:compression_error}.
The advantage of this method cannot be easily shown for \(N = 14\) because, taking \chicompete{} as an upper bound for dynamic growth of the bond size, the \(\chi\) required for \(\mathcal{I} = \num{e-4}\) is already close to \chicompete{}.
Truncation much lower than \chicompete{} results in an unacceptable accumulation of MPS truncation error.
However, \Cref{fig:reynolds_scaling} shows that the room for dynamic growth of \(\chi\) increases with the size of the system, as shown by the difference between \chicompete{} and \(\chi\).
Thus, a significant advantage over the DNS representation is achievable for large \(N\).
For example, at \(\Rey{} = 8 \times 10^{4}\) (\(r = 4096\)) and considering \(\rho\), the MPS truncation ratio is 1:11,715.
\section{Conclusions}\label{sec:conclusions}
A matrix product states (MPS) time evolution algorithm is developed for tensor network (TN) simulation of a coupled non-linear system of six PDEs describing the transport of  a two-dimensional unsteady compressible reacting shear flow.
An MPS algorithm, mirroring the MacCormack's finite difference discretizations method, yields excellent agreement with direct numerical simulation (DNS)  while using less than half its degrees of freedom.
Equivalent under-resolved DNS (URDNS) simulations cannot attain the same agreement, thus demonstrating the key advantage of MPS in using the available dof more efficiently.
The MPS calculations also faithfully capture important phenomenological aspects of compressible reacting flows, such as the reduction in mixing due to compressibility, heat release, and formation of eddy shocklets at high Mach numbers.
Analysis of DNS is performed and supports the conclusion that MPS truncation error decreases with increasing system size \(N\) because the required \(\chi\) grows at worst \ord{\Rey^{0.002}} for this flow.
Therefore, as \(N\) grows the magnitude of truncated singular values (i.e., the error) will become increasingly small.
It is also shown that MPS truncation error can be reduced if the number of operations per timestep are decreased.
Therefore, the potential for compression ratios of \(K = \numrange{e-5}{e-1}\) is demonstrated as shown in \cref{fig:dof_vs_PT}.
Compression ratios in this order would directly address current scaling limitations  in  turbulent reacting flows involving hundreds of chemical species with  complex reactions~\cite{dimotakis2005TURBULENT,driscoll2020Premixed,vidal2008Class}.
These truncation ratios will manifest as a significant reduction in the computational resources and time required for complex reacting flows~\cite{wang2018Direct,yang2013Largeeddy,aitzhan2023PeleLMFDF,aitzhan2025Onthefly,aitzhan2025Reduceda}.
Obtaining this advantage  is the key to facilitating efficient computation of  complex combustion  processes such as  ignition, extinction, and flame instabilities~\cite{domingo2023Recent,wang2018Direct}.
Another potential direction for future work is to adapt the MPS algorithm for numerical methods with fewer operations per timestep, which would directly address the MPS truncation error as discussed in \cref{sec:compression_error}.
Such efforts would  allow the benefits of MPS to be realized at larger system sizes.

More complex TN ans\"atze such as multi-scale entanglement renormalization(MERA)~\cite{vidal2008Class,evenbly2009Algorithms} and projected entangled pair states (PEPS)~\cite{ran2020Tensora,orus2019Tensora,cirac2021Matrix} may offer more advantages in terms of scaling.
Lastly, it is to be noted that the MPS algorithm as implemented here can be directly ported to existing quantum computers~\cite{tennie2025Quantuma,alipanah2025Quantum,siegl2025TensorProgrammable,smith2024ConstantDepth,malz2024Preparation,ran2020Encoding,jaksch2023Variationala,pool2024Nonlinear}.
The motivation for extending to quantum computers is that \ord{\Rey^{0.002}} scaling could be further improved allowing for an exponential decrease in memory compared to classical implementations of MPS~\cite{lubasch2020Variational}.

\section{Acknowledgments}

The work at Pitt is sponsored by the U.S. Air Force Office of Scientific Research (AFOSR) under Grant FA9550-23-1-0014, and by the University of Pittsburgh Momentum Funds.
The work at Hamburg is supported by the European Union’s Horizon Program (HORIZON-CL42021-
DIGITALEMERGING-02-10) Grant Agreement 101080085 QCFD, and the Hamburg Quantum Computing Initiative (HQIC) project EFRE that is co-financed by ERDF of the European Union, and by the Funds of the Hamburg Ministry of Science, Research, Equalities and Districts (BWFGB). Computational support is provided by the Center for Research Computing of the University of Pittsburgh, RRID:SCR\_022735, through the H2P cluster supported by the NSF Award  OAC-2117681.
\clearpage
\appendix{}
\section{Non-Dimensionalization Factors}\label{app:non_dimen}
\Cref{eqn:cons_of_mass,eqn:cons_of_momentum,eqn:cons_of_energy,eqn:species} are non-dimensionalized by
\begin{align}
    x^*    & = \frac{x}{L} \quad  y^* = \frac{y}{L} \quad u^* = \frac{u}{U_o} \quad v^* = \frac{v}{U_o}                  ,                           \\
    \rho^* & = \frac{\rho}{\rho_o} \quad  p^* = \frac{p}{\rho_oU_o^2} \quad e_T^* = \frac{e_T}{U_o^2} \quad T^* = \frac{T}{T_o} ,                    \\
    t^*    & = \frac{t}{\mathcal{T}} \quad \mathcal{T} = \frac{L}{U_o}  \quad q^* = \frac{q}{\rho_o U_o^3} \quad \tau^* = \frac{\tau}{\rho_o U_o^2}, \\
    \Rey{} & = \frac{U_o L}{\nu} \quad \Ma{} = \frac{U_o}{a_o} \quad \Da{} = \frac{\kappa L}{U_o} \quad \Pe{}  = \frac{U_o L}{\Gamma},               \\
    a_o    & = \sqrt{\gamma \mathcal{R} T_o}
\end{align}
Here \(U_o\) is the free stream velocity, \(\nu\) is the kinematic viscosity, \(\Gamma\) is the mass diffusion coefficient, \(a\) is the speed of sound, \(L\) is the domain size, \(\kappa\) is the rate of reaction, \(\tau\) denotes the viscous shear stresses, and \(\mathcal{R}\) is the specific gas constant.
Subscript \(o\) denotes free stream quantities at time \(t=0\).
In the remainder of this section the asterisks denoting a non-dimensionalized quantity are dropped.
The governing equations can be arranged into vector form as shown in \cref{eqn:gov}.
To begin, the \(\vec{U}\) component collects the time derivative terms as
\begin{equation}
    \pfrac{\vec{U}}{t} = \pfrac{}{t}\begin{bmatrix}\rho \\\rho u\\\rho v\\ \rho e_T\\\rho c_1\\ \rho c_2\end{bmatrix} = \pfrac{}{t}\begin{bmatrix}U_1 \\U_2\\U_3\\U_4\\U_5\\U_6\end{bmatrix}.\label{eqn:U}
\end{equation}
The \xx{} derivative terms are similarly collected in \(\vec{F}\) in the form
\begin{equation}
    \pfrac{\vec{F}}{x} =\pfrac{}{x}\begin{bmatrix}\rho u                                           \\[5pt]
        \rho u^2 + p -  \tau_{xx}                        \\[5pt]
        \rho uv - \tau_{xy}                              \\[5pt]
        u(\rho e_T + p)  + q_x - u\tau_{xx} - v\tau_{xy} \\[5pt]
        \rho c_1 u - \frac{1}{\Pe{}}\pfrac{\rho c_1}{x}  \\[5pt]
        \rho c_2 u - \frac{1}{\Pe{}}\pfrac{\rho c_2}{x}
    \end{bmatrix}
    = \pfrac{}{x}\begin{bmatrix}F_1\\[5pt]F_2\\[5pt]F_3\\[5pt]F_4\\[5pt]F_5\\[5pt]F_6\end{bmatrix},\label{eqn:F}
\end{equation}
and the \yy{} terms are collected in \(\vec{G}\) as
\begin{equation}
    \pfrac{\vec{G}}{y} =\pfrac{}{y}\begin{bmatrix}\rho v                                          \\[5pt]
        \rho uv -  \tau_{yx}                            \\[5pt]
        \rho v^2 + p - \tau_{yy}                        \\[5pt]
        v(\rho e_T + p) + q_y - v\tau_{yy} - u\tau_{yx} \\[5pt]
        \rho c_1 v - \frac{1}{\Pe{}}\pfrac{\rho c_1}{y} \\[5pt]
        \rho c_2 v - \frac{1}{\Pe{}}\pfrac{\rho c_2}{y}
    \end{bmatrix}
    = \pfrac{}{y}\begin{bmatrix}G_1\\[5pt]G_2\\[5pt]G_3\\[5pt]G_4\\[5pt]G_5\\[5pt]G_6\end{bmatrix}.\label{eqn:G}
\end{equation}
Lastly, the chemical source terms are placed in \(\vec{S}\) in the form
\begin{equation}
    \vec{S} = \begin{bmatrix}0 & 0 & 0 & 0 & -\Da{}\rho c_1 c_2 & -\Da{}\rho c_1 c_2\end{bmatrix}^{\text{T}},\label{eqn:S}
\end{equation}
where all components are zero except those corresponding to species conservation.
The non-dimensionalized components of the heat flux\footnote{No distinction is made between \Peclet{} for mass and thermal diffusion because Lewis number is taken as unity in all simulations.} and stress tensor are
\begin{align}
    \tau_{xx} & = \frac{1}{\Rey{}}\left(-\frac{2}{3} (\nabla\cdot\vec{V}) + 2  \frac{\partial u}{\partial x}\right),                       \\
    \tau_{yy} & = \frac{1}{\Rey{}}\left( -\frac{2}{3} (\nabla\cdot\vec{V}) + 2 \frac{\partial v}{\partial y}\right),                       \\
    \tau_{yx} & =  \frac{1}{\Rey{}}\left( \frac{\partial u}{\partial y} + \frac{\partial v}{\partial x}   \right),\label{eqn:shear_stress} \\
    q_x       & = - \frac{1}{(\gamma -1)\MA{}\Pe{}} \frac{\partial T}{\partial x},                                                         \\
    q_y       & =  - \frac{1}{(\gamma -1)\MA{}\Pe{}} \frac{\partial T}{\partial y}.\label{eqn:heat_flux}
\end{align}
The system is closed for \(p\) and \(T\) by the equation of state and assuming a calorically perfect gas.
\section{Initial Perturbation}\label{app:perturbation}
A perturbation is added to the initial velocity fields \(u(x,y,0)\) and \(v(x,y,0)\) to start the characteristic Kelvin-Helmholtz instability \cite{drazin2002Introduction}.
The perturbation has the form
\begin{align}
    \begin{split}
        N_x(x,y) & = \frac{2L}{\delta_i^2}\Bigg( \sin{\left(\frac{8\pi x}{L}\right)} + \sin{\left(\frac{24\pi x}{L}\right)} + \sin{\left(\frac{6\pi x}{L}\right)} \Bigg)\\
        &\times\Bigg( (y-y_{\text{min}}) \exp{\left(-\left(\frac{y-y_{\text{min}}}{\delta_i}\right)^2\right)}\\
        &+ (y-y_{\text{max}}) \exp{{\left(-\left(\frac{y-y_{\text{max}}}{\delta_i}\right)^2\right)}}\Bigg),\end{split}              \label{eqn:Noise_x} \\
    \begin{split}
        N_y(x,y) & = \pi \Bigg(6\cos{\left(\frac{6\pi x}{L}\right)} \Bigg) \\
        &\times\Bigg(  \exp{\left(-\left(\frac{y-y_{\text{min}}}{\delta_i}\right)^2\right)} +
        \exp{{\left(-\left(\frac{y-y_{\text{max}}}{\delta_i}\right)^2\right)}}\Bigg),
    \end{split}\label{eqn:Noise_y}
\end{align}
where \(y_{\text{min}} = 0.35\), \(y_{\text{max}} = 0.55\), and \(\delta_i = 3 \Delta y\).
The perturbation fields \(N_x\) and \(N_y\) are normalized by \(\sqrt{N_x^2 + N_y^2}\) so they are between 0 and 1 and then scaled by a factor of \(U_o/\mathcal{F}\).
The factor \(\mathcal{F}\) was equal to 40 so that the maximum amplitude of perturbation  was at most 2.5\% of \(U_o\).
\section{Timestep Size}\label{app:tstep}
The timestep size is calculated once at the beginning of the simulation as~\cite{hoffmann2004Computational}
\begin{equation}
    \Delta t = \left(\frac{\sigma}{1 + \frac{2}{\Rey_\Delta}}\right)\left( \frac{1}{\frac{U_o}{\Delta x} + + a_o \sqrt{\frac{1}{\Delta x^2} + \frac{1}{\Delta y^2}}}\right),\label{eqn:dt}
\end{equation}
where \(\Delta x\) and \(\Delta y\) are the grid spacing, \(\sigma\) is a safety factor, and \(\Rey{}_\Delta\) is the minimum cell Reynolds number equal to
\begin{equation}
    \Rey{}_\Delta = \min{}\left( \frac{\Delta x}{L} \Rey{} , \frac{\Delta y}{L} \Rey{} \right).
\end{equation}
\section{MacCormack's Method}\label{app:MacCormacks}
MacCormack's method is a predictor-corrector technique.
The steps of MacCormack's method will be illustrated in terms of the solution vector given by \cref{eqn:gov} where each component \(k\) is denoted by \((U_k)^n_{i,j}\).
Here \(i,j\) correspond to the spatial indices and \(n\) corresponds to the current timestep.
The predictor step begins by computing
\begin{multline}
    \pfrac{(U_k)_{i,j}^n}{t} = - \Bigg( \frac{(U_k)_{i+1,j}^n - (U_k)_{i,j}^n}{\Delta x}\\ + \frac{(U_k)_{i,j+1}^n - (U_k)_{i,j}^n}{\Delta y}  \Bigg),
\end{multline}
where the spatial derivatives have been evaluated using forward differences and \cref{eqn:F,eqn:G}.
The predictor value of \(\overline{U}_k\) at time \(n+1\) is then calculated by
\begin{equation}
    (\overline{U}_k)_{i,j}^{n+1} = (U_k)^n_{i,j} + \Delta t \pfrac{(U_k)_{i,j}^n}{t}.
\end{equation}
The corrector step requires evaluating \cref{eqn:F,eqn:G} which necessitates calculation of the predictor step primitive variables.
The primitive variables are recovered by elementwise (Hadamard) division according to \cref{eqn:U} as (shown only for \(u\) as an example)
\begin{equation}
    (\overline{u})_{i,j}^{n+1} = \frac{(\overline{U}_2)_{i,j}^{n+1}}{(\overline{U}_1)_{i,j}^{n+1}}.
\end{equation}
The corrector step derivative (denoted by the asterisk) is then calculated as
\begin{multline}
    \pfrac{(U_k)_{i,j}^*}{t} = - \Bigg( \frac{(\overline{U}_k)_{i,j}^{n+1} - (\overline{U}_k)_{i-1,j}^{n+1}}{\Delta x}\\ + \frac{(\overline{U}_k)_{i,j}^{n+1} - (\overline{U}_k)_{i,j-1}^{n+1}}{\Delta y}  \Bigg),
\end{multline}
where spatial derivatives have been evaluated using backward differences.
The average rate of change can then be calculated from the predictor and corrector rates of change as
\begin{equation}
    \pfrac{(U_k)_{i,j}^{\text{Avg}}}{t} = \frac{1}{2}\left(\pfrac{(U_k)_{i,j}^n}{t} + \pfrac{(U_k)_{i,j}^*}{t}\right).
\end{equation}
The corrector value of \(U_k\) is then obtained as
\begin{equation}
    (U_k)_{i,j}^{n+1} = (U_k)_{i,j}^n + \Delta t \pfrac{(U_k)_{i,j}^{\text{Avg}}}{t}.
\end{equation}
Lastly, the corrector values of the primitive variables are recovered in preparation for the next timestep (again using \(u\) as an example) as
\begin{equation}
    (u)_{i,j}^{n+1} = \frac{(U_2)_{i,j}^{n+1}}{(U_1)_{i,j}^{n+1}}.
\end{equation}
\section{Index Ordering}\label{app:indexing}
\begin{figure}
    \includegraphics{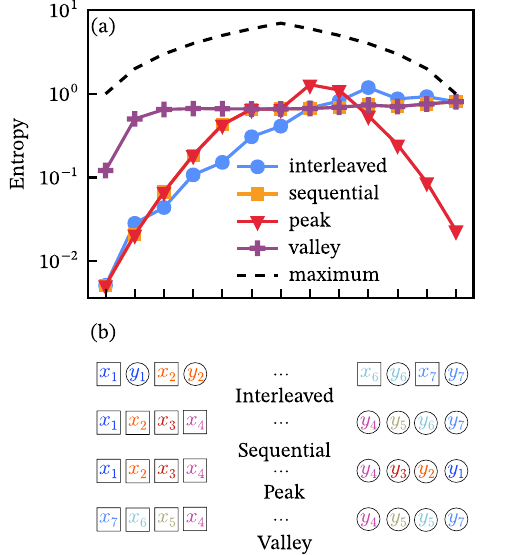}
    \caption{The entropy \(S\) for each bond for the four indexing schemes investigated. The black dashed line is the maximum entropy equal to \(S_{\text{max}} = \log_2{(d^n)}\) where \(d\) is the physical index size (equal to 2) and \(n\) is the bond number~\cite{wilde2016Classical}.}\label{fig:index_ordering}
\end{figure}

For two-dimensional Cartesian grids the physical indices are labeled as \(x_1 x_2 \ldots x_N\) and \(y_1 y_2 \ldots y_N\) to distinguish the \(x\) and \(y\) coordinate directions.
The arrangement of the indices impacts the MPS truncation error.
Four possible choices of indexing are investigated and shown as
\begin{equation}\label{eqn:index_options}
    \begin{gathered}
        \underbrace{x_1 y_1 x_2 y_2 \ldots x_N y_N}_{\text{interleaved}} \qquad \underbrace{x_1 x_2 \ldots x_N y_1 y_2 \ldots y_N}_{\text{sequential}}\\
        \underbrace{x_1 \ldots x_{N-1} x_N y_N y_{N-1}\ldots y_1}_{\text{peak}} \qquad \underbrace{x_N x_{N-1} \ldots x_1 y_1 \ldots y_{N-1} y_N}_{\text{valley}}
    \end{gathered}
\end{equation}
Indices corresponding to the smallest and largest scales are \(x_1,y_1\) and \(x_N,y_N\), respectively.
The large scales in the TDJ are the most strongly correlated.
This provides a motivation to place the large scales near the middle of the MPS\@.
\Cref{fig:index_ordering} illustrates the impact of the indexing schema in terms of entropy which is a measure of correlation between scales and is defined for at given bond as~\cite{wilde2016Classical}
\begin{equation}
    S = - \frac{1}{E} \sum_{i=1}^n \lambda_i^2 \log_2{\left(\frac{\lambda_i^2}{E}\right)} \quad \text{with} \quad E = \sum_{i=1}^n \lambda_i^2,\label{eqn:entropy}
\end{equation}
where \(\lambda_i\) are the singular values (see \cref{eqn:schmdit_decomp}).
Stronger correlations among the physical scales should correspond to larger entropies.
Therefore, a good indexing scheme will have large entropies at the large scales and minimize the entropy at all other scales as reflected for peak in \cref{fig:index_ordering}.
The remaining schemes do not exploit the structure in the correlations of the TDJ and correspondingly the entropy is large even at smaller scales.
\section{Finite Difference MPO Construction}\label{app:MPO_construction}
The purpose of this Appendix is to illustrate the process of constructing MPOs for finite differences such as
\begin{gather}
    f'_i = \frac{f_{i+1} - f_{i-1}}{2 \Delta x},\label{eqn:diff_stencil}
\end{gather}
where the MPO is exact and has the minimal possible bond dimension \(C\).
As will be shown, this process consists of defining left and right operators from which the desired finite difference can be assembled.

\begin{figure}[b]
    \includegraphics{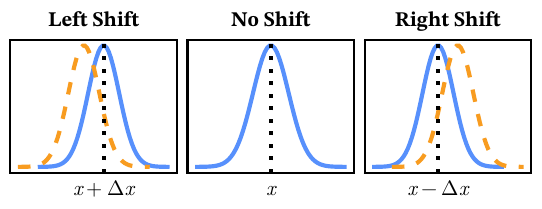}
    \caption{Illustration of the impact on \xx{} from the LS and RS operators with a shift of \(\Delta x\). After RS the point that was previously \xx{} has become \(x - \Delta x\). After the LS the point that was previously \xx{} has become \(x + \Delta x\).}\label{fig:left_and_right_shift}
\end{figure}

Consider a left shift operator \LS{} which shifts all points in the domain one grid point leftwards.
For an eight point grid \LS{} is written as
\begin{gather}
    \LS{} = \begin{bmatrix}
        0 & 1 & 0 & 0 & 0 & 0 & 0 & 0 \\
        0 & 0 & 1 & 0 & 0 & 0 & 0 & 0 \\
        0 & 0 & 0 & 1 & 0 & 0 & 0 & 0 \\
        0 & 0 & 0 & 0 & 1 & 0 & 0 & 0 \\
        0 & 0 & 0 & 0 & 0 & 1 & 0 & 0 \\
        0 & 0 & 0 & 0 & 0 & 0 & 1 & 0 \\
        0 & 0 & 0 & 0 & 0 & 0 & 0 & 1 \\
        1 & 0 & 0 & 0 & 0 & 0 & 0 & 0
    \end{bmatrix}.\label{eqn:LS}
\end{gather}
The action of the \LS{} operator is illustrated for a one-dimensional grid in \cref{tab:shift_and_encoding}\footnote{This encoding has assumed column-major ordering where the least significant bit (fastest varying) is the first bit. If row-major ordering had been chosen the least significant bit would be the last bit and the value of \(f\) at \(1\) would have been indexed by \(001\).}.
\begin{table}
    \caption{Impact of \LS{} operator on \(2^3\) grid from 0 to 7 in terms of binary encoding of indices. For example, \(f_0\) is at grid position 0 in the unshifted grid and on the left shifted grid is at position 7.}\label{tab:shift_and_encoding}
    \begin{tabular}{c}
        \toprule
        \(f\)   \\
        \midrule
        \(f_0\) \\
        \(f_1\) \\
        \(f_2\) \\
        \(f_3\) \\
        \(f_4\) \\
        \(f_5\) \\
        \(f_6\) \\
        \(f_7\) \\
        \bottomrule
    \end{tabular}
    \begin{tabular}{c c}
        \toprule
        \xx{} & \(\xx{}_{i_1i_2i_3}\) \\
        \midrule
        0     & 000                   \\
        1     & 100                   \\
        2     & 010                   \\
        3     & 110                   \\
        4     & 001                   \\
        5     & 101                   \\
        6     & 011                   \\
        7     & 111                   \\
        \bottomrule
    \end{tabular}
    \begin{tabular}{c c}
        \toprule
        \(LS(\xx{})\) & \(LS(\xx{}_{i_1i_2i_3})\) \\
        \midrule
        7             & 111                       \\
        0             & 000                       \\
        1             & 100                       \\
        2             & 010                       \\
        3             & 110                       \\
        4             & 001                       \\
        5             & 101                       \\
        6             & 011                       \\
        \bottomrule
    \end{tabular}
\end{table}
A right shift operator \RS{} can be similarly formulated.
The action of \LS{} and \RS{} operators is illustrated in \cref{fig:left_and_right_shift}.
These operator matrices can be formed into an MPO by appropriately reshaping and permuting the input and output indices (columns and rows, respectively) as
\begin{gather}
    I_1I_2I_3O_1O_2O_3 \rightarrow I_1O_1I_2O_2I_3O_3,\label{eqn:MPO_permuted}
\end{gather}
where the indices of the rows and columns are denoted by \(O\) and \(I\), respectively.
A series of SVDs is performed to convert the matrix into an MPO as was done for the MPS in \cref{sec:MPS_Description}.
Each pairing \(I_iO_i\) corresponds to a single combined index with \(\sizeof{I_iO_i} = 4\).
This simple approach scales exponentially with system size and therefore must be truncated.

\begin{figure}
    \includegraphics{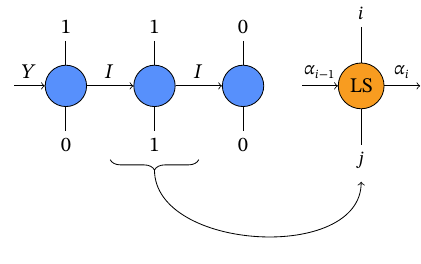}
    \caption{\LS{} MPO as a string of interconnected FSM where each tensor is an FSM\@. Instructions are passed to the next FSM via the bond indices which are shown explicitly for the middle tensor.}\label{fig:LS_FSM}
\end{figure}

The methodology of Ref.~\cite{crosswhite2008Finite} offers a more efficient and computationally feasible approach.
It consists of defining a finite state machine (FSM) that produces the bulk tensor of the desired MPO\@.
FSM are a way to describe a reactive system characterized by states.
Familiar examples include traffic lights and elevators.
In the case of a traffic light, there are three states: green, yellow, and red.
The next state of the FSM depends on the current state.
If the light is green, its next state will be yellow.
Therefore, green to yellow is an allowed transition of the FSM while
yellow to green is not.
In addition to states, the FSM may have an input and output value associated with a transition.
Consider again the action of the \LS{} operator shown in \cref{tab:shift_and_encoding}.
For the \LS{} MPO, consider each individual tensor as a FSM so that the MPO is composed of an interconnected string of FSMs as shown in \cref{fig:LS_FSM}.
Instructions from one FSM can be passed to the adjacent FSM via the bonds.
Now consider a single bulk (interior) tensor or FSM\@.
The bond indices on the left and right of the FSM correspond to the incoming (from previous FSM) and outgoing (to next FSM) instructions.
The input and output values of the FSM correspond to the physical indices (\(i\) and \(j\)) of the MPO, respectively.
As will become clear, the FSM of the \LS{} operator requires two possible states: identity (\II{}) and binary subtraction (\YY{}).
Binary subtraction is necessary because the input/output values of the MPO are encoded in binary as described in \cref{sec:MPS_Description}.
The following mapping is adopted to simplify notation: \(\II{} \rightarrow 0\) and \(\YY{} \rightarrow 1\).
Considering the FSM shown in the bottom half of \cref{fig:LS_FSM}, a table of allowed transitions can be constructed as:
\begin{itemize}
    \item Input instruction (\(\alpha_{i-1}\)) is \II{} with an FSM input value of \(i=0\). The identity instruction \II{} returns as output the same input so that \(j = 0\). An output value of 0 has zero extra bits to carry forward to the next FSM, therefore the instruction to the next machine (\(\alpha_i\)) is 0 which corresponds to \II{}.
    \item Input instruction (\(\alpha_{i-1}\)) is \II{} with an FSM input value of \(i=1\). The identity instruction \II{} returns as output the same input so that \(j=1\). There are zero bits to carry forward, therefore the instruction to the next machine (\(\alpha_i\)) is \II{}.
    \item Input instruction (\(\alpha_{i-1}\)) is \YY{} with an FSM input value of \(i=1\). The binary subtraction instruction \YY{} with an input value of 1 (01 in binary) results in an output value \(j = 0\). There are zero bits to carry forward, therefore the instruction to the next machine (\(\alpha_i\)) is \II{}.
    \item Input instruction  (\(\alpha_{i-1}\)) is \YY{} with an FSM input value of \(i=0\). The binary subtraction instruction \YY{} with an input value of 0 results in 11 in binary (assuming two's complement). The output value is then \(j = 1\) with an extra bit left over which is carried forward to the next machine by \(\alpha_i\) as an instruction of \YY{}.
\end{itemize}
These transitions are summarized in \cref{tab:FSM_one_site} and can be used to define an array of size \(C \times d \times d \times C\) which is the bulk tensor of the \LS{} MPO\@.
A one is placed at each element that corresponds to an allowed transition.
All other elements are zero reflecting transitions that are not allowed.
\begin{table}
    \caption{Single FSM (tensor) encoding of \cref{fig:LS_FSM}.}\label{tab:FSM_one_site}
    \begin{tabular}{c c c c}
        \toprule
        \(\alpha_{i-1}\) & \(i\) & \(j\) & \(\alpha_{i}\) \\
        \midrule
        \II{}            & 0     & 0     & \II{}          \\
        \II{}            & 1     & 1     & \II{}          \\
        \YY{}            & 1     & 0     & \II{}          \\
        \YY{}            & 0     & 1     & \YY{}          \\
        \bottomrule
    \end{tabular}
\end{table}
The \LS{} MPO has a maximum bond dimension of \(C=2\) because it requires only two states to describe it: identity and binary subtraction.

All that remains to specify for the MPO are the left and right boundary tensors.
These tensors can be created by contracting the bulk tensor with a specially formulated \textit{terminator} tensor.
The left terminator is a one-dimensional array that encodes the input instruction to the FSM\@.
Each element of the terminator corresponds to an instruction (state) of the FSM\@.
In the case of the \LS{} MPO the left terminator is a \(1 \times C\) array where the first and second elements correspond to the \II{} and \YY{} instructions, respectively.
The input instruction to the FSM begins with state \YY{} so the left terminator should be formulated as \cref{eqn:left_terminator}.
This can be contracted with a bulk tensor over index \(\alpha_{i-1}\) to produce the left boundary tensor with indices \(i\), \(j\), and \(\alpha_i\).
\begin{gather}
    \text{Left Terminator} = \begin{bmatrix}
        0 & 1
    \end{bmatrix}\label{eqn:left_terminator}
\end{gather}
The right terminator encodes the boundary conditions of the \LS{} operator.
It is a \(C \times 1\) array and similar to the left terminator, the elements correspond to the instructions of the FSM\@.
In the case of the \LS{} MPO the right terminator is written as \cref{eqn:right_terminator} which yields a transformation as \cref{eqn:periodic_boundary}.
Just as with the left boundary, the right terminator is contracted with a bulk tensor over index \(\alpha_{i}\) which yields the right boundary tensor with indices \(\alpha_{i-1}\), \(i\), and \(j\).
\begin{gather}
    \text{Right Terminator} = \begin{bmatrix}
        1 \\
        1
    \end{bmatrix}\label{eqn:right_terminator}\\
    \begin{bmatrix}
        f_{1} & f_{2} & \cdots & f_7 & f_{8}
    \end{bmatrix} \rightarrow
    \begin{bmatrix}
        f_{2} & f_{3} & \cdots & f_8 & f_1
    \end{bmatrix}\label{eqn:periodic_boundary}
\end{gather}
Open boundary conditions could be implemented by changing the second element (corresponding to state \YY{}) to a 0 which would yield a transformation as \cref{eqn:open_boundary}.
\begin{gather}
    \begin{bmatrix}
        f_{1} & f_{2} & \cdots & f_7 & f_{8}
    \end{bmatrix} \rightarrow
    \begin{bmatrix}
        f_{2} & f_{3} & \cdots & f_8 & 0
    \end{bmatrix}\label{eqn:open_boundary}
\end{gather}
Everything needed to manually construct the \LS{} MPO with minimal bond dimension is now known.
The \RS{} MPOs is formulated similarly with the only difference being that the state \YY{} is replaced with binary addition.\footnote{The formulation shown here assumes column-major ordering, and in row-major ordering the \ LS {} would correspond to binary addition and RS would correspond to binary subtraction.}
The \LS{}, \RS{}, and identity MPOs provide the basic building blocks from which finite difference operators can be constructed.

To illustrate this process, consider the central difference for the first derivative given in \cref{eqn:diff_stencil}.
The MPO is formulated by combining the \LS{} and \RS{} operators.
The \RS{} MPO corresponds to \(f_{i-1}\) and the \LS{} MPO corresponds to \(f_{i+1}\) as shown in \cref{fig:left_and_right_shift}.
Each \LS{} and \RS{} MPO is scaled by a finite difference coefficient of \(\pm 1/2\Delta x\) depending on the term in \cref{eqn:diff_stencil}.
The coefficient is applied via the left terminator by multiplying all elements of the terminator by the coefficient and then contracting the terminator with a bulk tensor.
The resulting central difference MPO is then calculated as
\begin{equation}
    \widehat{\partial}_{CD} = LS \oplus RS,\label{eqn:central_diff_MPO}
\end{equation}
where \(\oplus\) is the direct sum.
The central difference operation requires three operations: binary subtract (\LS{}), binary add (\RS{}), and identity.
Therefore, it has \(C = 3\) and the resulting MPO should be truncated accordingly after performing the direct sum.
More complicated finite difference stencils require multiple shifts (e.g., \(f_{i+4}\)).
This can be achieved by repeatedly contracting RS or LS MPOs together.
Care should be taken to ensure the finite difference coefficients are not applied until after the shift is complete.
\section{Hadamard Product}\label{app:hadamard_mult}
The Hadamard product (elementwise product) is a key operation in the solution of non-linear PDEs.
In terms of MPS, the Hadamard product \mpsC{} of \mpsA{} and \mpsB{} begins by first doubling the physical indices of one of the input MPS at the cost of \ord{N\chi^2d^3}.
This effectively converts the MPS into an MPO, but more importantly with \(C = \chi\).
The doubling of physical indices is done using the identity tensor \(\delta_{i\,i}^{i}\), so we define the MPOs
\begin{align}
    \begin{split}
        \widehat{\Phi}_A & = \sum_{\mathclap{\substack{i_1,i_2,i_3 \\\alpha_1,\alpha_2}}} (A_{\alpha_1}^{i_1}\delta_{i_1 i_1}^{i_1}) (A_{\alpha_1,\alpha_2}^{i_2}\delta_{i_2 i_2}^{i_2}) (A_{\alpha_3}^{i_3}\delta_{i_3 i_3}^{i_3}) \ket{i_1i_2i_3} \bra{i_1i_2i_3},\\
        & = \sum_{\mathclap{\substack{i_1,i_2,i_3 \\\alpha_1,\alpha_2}}} A_{\alpha_1}^{i_1,i_1}A_{\alpha_1,\alpha_2}^{i_2,i_2}A_{\alpha_3}^{i_3,i_3}\ket{i_1i_2i_3} \bra{i_1i_2i_3},
    \end{split}\label{eqn:hadamard}
\end{align}
where \(A\) denote the individual tensors of \mpsA{}.
The MPO \(\widehat{\Phi}_A\) is then applied to \mpsB{} in the manner described in \cref{sec:MPO} in the form
\begin{equation}
    \mpsC{} = \mpsA{}\odot\mpsB{} =  \widehat{\Phi}_A \mpsB{},\label{eqn:hadamard_product}
\end{equation}
where \(\odot\) denotes the Hadamard product.
The process is analogous to the Hadamard product of two vectors wherein one of the vectors is distributed along the diagonal of a matrix as shown in
\begin{equation}
    \vec{a} \odot \vec{b} =
    \begin{bmatrix}
        a_1 \\ a_2 \\ \vdots \\ a_n
    \end{bmatrix}
    \odot
    \begin{bmatrix}
        b_1 \\ b_2 \\ \vdots \\ b_n
    \end{bmatrix} =
    \underbrace{\begin{bmatrix}
            a_1    & 0      & \cdots & 0      \\
            0      & a_2    & \ddots & \vdots \\
            \vdots & \ddots & \ddots & 0      \\
            0      & \cdots & 0      & a_n
        \end{bmatrix}}_{A}
    \begin{bmatrix}
        b_1 \\ b_2 \\ \vdots \\ b_n
    \end{bmatrix} =
    \begin{bmatrix}
        a_1b_1 \\ a_2b_2 \\ \vdots \\ {a_n}{b_n}
    \end{bmatrix},
\end{equation}
where the matrix \(A\) corresponds to \(\widehat{\Phi}_A\) in \cref{eqn:hadamard_product}.
The MPO-MPS product scales as \ord{NC^3\chi^3}.
The MPO bond indices are now of size \(\chi^2\), resulting in a nominal scaling of \ord{\chi^6} if the SVD sweeps to compress back to \(\chi\) are included in the cost.
As stated in \cref{sec:MPS_Operations}, the fit algorithm is  used here, which results in a Hadamard scaling of \ord{\chi^4}.
See also Ref.~\cite{michailidis2025Elementwise} for an alternative formulation of the Hadamard product.
\section{Hadamard Division}\label{app:hadamard_div}
An operation unique to this work is the Hadamard division (elementwise division).
As shown in \cref{fig:alg} the primitive variables have to be recovered and updated twice per timestep.
This is analogous to recovering, for example, the velocity field \(u = U_2 \oslash \rho\) where \(\oslash\) denotes the Hadamard division.
In the MPS ansatz this operation is carried out in two steps.
To illustrate the process, assume that \(\mpsC{} = \mpsA{} \oslash \mpsB{}\) is desired.
First, the inverse of \mpsB{} is calculated.
Then, the Hadamard product of \mpsA{} and \(\ket{\psi_B^{-1}}\) is taken.
Converting \mpsB{} to an MPO as
\begin{equation}
    \mpsB{} \rightarrow \widehat{\Psi}_B,
\end{equation}
is done by doubling the physical indices at a cost of \ord{N\rchi^2d^3}, as in \cref{eqn:hadamard}.
Next, a cost function \(E\) is formulated as
\begin{align}
    \begin{split}
        E & = ||\widehat{\Psi}_B\ket{x} - \ket{\mathbb{1}}||_2^2,                                                                                                                                             \\
        & = \braket{x|\widehat{\Psi}_B^\dagger \widehat{\Psi}_B | x} - \braket{x |\widehat{\Psi}_B^\dagger | \mathbb{1}} - \braket{\mathbb{1} |\widehat{\Psi}_B | x} + \braket{\mathbb{1}|\mathbb{1}},
    \end{split}\label{eqn:cost_function}
\end{align}
where \(\ket{x}\) is an initial guess MPS (e.g., a random MPS) and \(\ket{\mathbb{1}}\) corresponds to the MPS representation of a vector with all elements equal to 1.
Considering the elements of \(\ket{x}\) as the parameters of \(E\), this represents a high dimensional highly non-linear optimization problem.
It can be made manageable using the strategy described in Ref.~\cite{schollwoeck2011densitymatrix}.
In brief, only the elements of a single tensor \(j\) of \(\ket{x}\) are optimized at a time so that the derivative of the cost function \(E\) becomes
\begin{equation}
    \pfrac{E}{\bra{x_j}} = \braket{x_{/j}|\widehat{\Psi}_B^\dagger \widehat{\Psi}_B | x} - \braket{x_{/j}|\widehat{\Psi}_B^\dagger |\mathbb{1}}\label{eqn:dEdx}
\end{equation}
where \(\bra{x_{/j}}\) represents an MPS with tensor \(j\).
A single optimization step of \(\ket{x}\) consists of iterating over the tensors of the MPS one at a time from left to right and right back to left.
The minimization at each individual tensor \(j\) is performed using \cref{eqn:dEdx} and the generalized minimal residual method (GMRES)~\cite{saad1986GMRES}.
A detailed description of the implementation is given in Ref.~\cite{catarina2023Densitymatrixa} in the context of the density matrix renormalization group (DMRG)~\cite{schollwock2011densitymatrix}.
The DMRG process is identical to the one described here except at each tensor an eigenvalue problem is solved using Lanczos~\cite{lanczos1950iteration} or Arnoldi~\cite{arnoldi1951principle} instead of an optimization problem using GMRES\@.

From \cref{eqn:cost_function} and the optimization routine described, the inverse of \mpsB{} is found as
\begin{equation}
    \ket{\psi_B^{-1}} = \argmin{(E)}.
\end{equation}
The contractions in the \(\widehat{\Psi}_B^\dagger \widehat{\Psi}_B\) term of \cref{eqn:dEdx} are the most expensive to perform and can be done at a cost of \ord{\chi^4} using the fit algorithm.
The process yields a new MPO with bond dimension of only \(\chi\).
The cost of performing GMRES at each local tensor is \ord{\chi^4}.
Therefore, the overall cost of finding \(\ket{\psi_B^{-1}}\) is \ord{\chi^4}.
The Hadamard division is completed by performing a Hadamard product as shown in
\begin{equation}
    \mpsC{} = \mpsA{} \odot \ket{\psi_B^{-1}},
\end{equation}
at a cost of \ord{\chi^4} using the fit algorithm.
Thus, Hadamard division has an overall scaling of \ord{\chi^4}.
\clearpage
\bibliography{references}
\end{document}